\documentclass[12pt,aps,showpacs]{revtex4}
\usepackage{amssymb,amsfonts,epsfig}
\newcommand{\beeq}{\begin{equation}}
\newcommand{\eneq}{\end{equation}}
\newcommand{\be}{\begin{eqnarray}}
\newcommand{\ee}{\end{eqnarray}}
\newcommand{\bpic}{\begin{picture}}
\newcommand{\epic}{\end{picture}}
\newcommand{\bs}{\begin{scriptsize}}
\newcommand{\es}{\end{scriptsize}}

\def\dd{\partial}
\def\la{\raise.16ex\hbox{$\langle$} \, }
\def\ra{\, \raise.16ex\hbox{$\rangle$} }

\def\a{\alpha}
\def\b{\beta}

\def\g{\gamma}

\def\D{\Delta}

\def\l{\lambda}

\def\Box{\kern1pt\vbox{\hrule height 1.2pt\hbox{\vrule width 1.2pt\hskip 3pt
   \vbox{\vskip 6pt}\hskip 3pt\vrule width 0.6pt}\hrule height 0.6pt}\kern1pt}
\def\gtwid{\mathrel{\raise.3ex\hbox{$>$\kern-.75em\lower1ex\hbox{$\sim$}}}}
\def\ltwid{\mathrel{\raise.3ex\hbox{$<$\kern-.75em\lower1ex\hbox{$\sim$}}}}
\begin{document}

\hspace{3cm}

\title{The Quantum Corrected Mode Function and Power Spectrum\\ for a Scalar Field during Inflation}

\author{V. K. Onemli}\email{onemli@itu.edu.tr}

\affiliation{$^{\ast}$Department of Physics, Istanbul Technical
University, Maslak, Istanbul 34469, Turkey}
\begin{abstract}
We compute the one- and two-loop corrected mode function of a
massless minimally coupled scalar endowed with a quartic
self-interaction in the locally de Sitter background of an
inflating universe for a state which is released in Bunch-Davies
vacuum at time $t=0$. We then employ it to correct the scalar's
tree-order scale invariant power spectrum $\Delta^2_\varphi$. The
corrections are secular, and have scale dependent part that can be
expanded in even powers of $k/(Ha)$, where $k$ is the comoving
wave number, $H$ is the expansion rate and $a$ is the cosmic scale
factor. At one-loop, the scale invariant shift in the power
spectrum grows as $(Ht)^2$ in leading order. The $k$-dependent
shifts, however, are constants for each mode, in the late time
limit. At two-loop order, on the other hand, the scale invariant
shift grows as $(Ht)^4$ whereas the $k$-dependent shifts grow as
$(Ht)^2$, in leading order. We finally calculate the scalar's
spectral index $n_\varphi$ and the running of the spectral index
$\alpha_\varphi$. They imply that the spectrum is slightly
red-tilted; hence, the amplitudes of fluctuations grow slightly
towards the larger~scales.
\end{abstract}


\maketitle \vskip 0.1in \vspace{.4cm}

\section{Introduction}
Inflationary expansion enhances quantum effects \cite{RWo1}. The
enhancement can be strong enough for quantum fields that are
classically conformally non-invariant and effectively massless, to
be realized even on cosmic scales. Massless minimally coupled
(MMC) scalars and gravitons are the two examples which possess
these attributes. The observed scalar perturbations \cite{MuCh}
and potentially observable tensor perturbations \cite{ASt1} are,
indeed, the amplified imprints of tiny quantum fluctuations of
inflatons (inflationary scalars) and gravitons on the cosmic
microwave anisotropy, respectively.

Thus, to study further possible quantum effects on cosmological
scales, we considered, in
Refs.~\cite{OnWo1,OnWo2,KaOnWo,BrOnWo,KaOn}, the MMC scalar field
with a quartic self-interaction $\lambda \varphi^4(x)$ in the
locally de Sitter spacetime. It is the cosmological constant
$\Lambda$ that drives inflation in the model. The constant
expansion rate $H\!=\!\sqrt{\Lambda/(D\!-\!1)}$ in $D$-spacetime
dimensions. The scalar $\varphi$ is a spectator to the
$\Lambda$-driven de Sitter inflation. We revealed several
intriguing quantum aspects of the model. The fully renormalized
vacuum expectation value of the energy density $\rho_{{\rm ren}}$
and pressure $p_{{\rm ren}}$ were computed
\cite{OnWo1,OnWo2,KaOnWo} using dimensional regularization at one
and two-loop orders. Although the classical covariant
stress-energy conservation law $\dot{\rho}=\!-(D-\!1)H(\rho+\!p)$
is obeyed by $\rho_{{\rm ren}}$ and $p_{{\rm ren}}$, the classical
weak energy condition (WEC) $\rho+\!p\!\geq0$ is violated on
cosmological scales---on average, not just in fluctuations. This
quantum anomaly is a two-loop effect. The parameter $w\equiv
p_{{\rm ren}}/\rho_{{\rm ren}}\!<\!-1$ at that order, thus, a
temporary phase of super-acceleration is induced as a quantum
effect. Physically, the effect can be described as follows.
Inflationary particle production implies the growth of the scalar
field strength. The scalar undergoes a random walk, so that, its
position in the quartic potential rises, on average. Hence, the
vacuum energy density increases: $\dot{\rho}_{{\rm ren}}\!>0$.
Combined with this fact, the stress-energy conservation implies
the violation of the WEC: $\rho_{{\rm ren}}\!+p_{{\rm ren}}\!<0$.
This process, however, is self-limiting. Because the field
develops \cite{BrOnWo} a positive self-mass squared as it grows,
the particle production is cut off. Moreover, as the scalar rises
up its potential the classical restoring force pushes it back
down. The field cannot continue rolling up its position and comes
to a halt. Hence, the effect in the system is self-terminated.

The stability of the system was studied in Ref.~\cite{KaOn}. We
putatively solved the linearized Schwinger-Keldish effective
scalar field equation in the late time limit $t\gg 0$, in leading
logarithm order, and obtained the mode function as \beeq\Phi(x,
\vec{k})\!\sim\!u(\infty, k) e^{i \vec{k}
\cdot\vec{x}}\Bigg\{\!1-\frac{\lambda}{2^4
3\pi^2}\!\left[\ln^2(a)\!-\!\frac{2\ln(a)}{3}\right]\!+\frac{\l^2}{2^8
3^3\pi^4}\!\left[\!\frac{11\ln^4(a)}{2}\!-\!\ln^3(a)\!\right]\!+\mathcal{O}(\l^3)\!\Bigg\}\;
.\label{oldmodesoln}\eneq Here $\vec{k}$ is the comoving
wavevector, $k$ is the comoving wave number and $a=e^{Ht}$ is
the~scale factor of the spacetime. The amplitude $u(\infty,
k)\!=\!H/\sqrt{2k^3}$ is the classical Bunch-Davies mode function
$u(t, k)$ in the late time limit $t\gg 0$.
Equation~(\ref{oldmodesoln}) implies that the perturbation breaks
down when $\ln(a)\!\sim~\!\!1/\sqrt{\l}$. For a coupling constant
$\lambda\ll 1$, one can have a long time evolution during which
the perturbation theory is still valid. As time evolves, the
amplitude is reduced. The linear perturbations do not grow in this
model. So, the model is stable.

Solution~(\ref{oldmodesoln}) for $\Phi(x, \vec{k})$ is correct in
leading logarithm order. However, due to the restrictions
\cite{KaOn} imposed on the solution, it neglects the sub-leading
terms and completely misses the wave number $k$-dependent shifts
that are essential in revealing the scale dependence of the
scalar's power spectrum. In de Sitter background, the one-loop
corrections to the mode functions in several different models have
been calculated, in leading logarithm order, under similar
restrictions. The one-loop corrections to the fermion and scalar
mode functions for a MMC scalar Yukawa coupled to a massless
fermion were calculated in Refs.~\cite{PrWo1} and \cite{DuWo},
respectively. The one-loop corrections to the photon and scalar
mode functions in MMC scalar quantum electrodynamics were
calculated in Refs.~\cite{PrToWo} and \cite{KaWo1}, respectively.
The one-loop correction to the fermion mode function for a
massless fermion coupled to gravity was calculated in
Ref.~\cite{MiWo2}. The one-loop correction to the scalar mode
function for a MMC scalar coupled to gravity was calculated in
Ref.~\cite{KaWo2}. As in Eq.~(\ref{oldmodesoln}) these solutions
were putative and neglected sub-leading terms. Except for the one
in Ref.~\cite{KaWo2}, they did not include the $k$-dependent
shifts.

A purpose of the present paper is to calculate the full one- and
the two-loop corrections to the classical Bunch-Davies scalar mode
function in our model. This is achieved by exactly solving the
linearized Schwinger-Keldish effective scalar field equation,
including the fully nonlocal and derivative corrections, by
applying the Green's function technique. In both loop orders, as
we show, the quantum corrections induce $k$-dependent shifts to
the scalar mode function. We also analyze how these shifts modify
the scalar's scale invariant tree-order power spectrum, its
spectral index and the running of the spectral index in this
paper.

The outline is as follows. In Sec.~\ref{sec:model} we specify the
background geometry, present the model, and briefly review the
computation of the tree-order (Bunch-Davies) scalar mode function.
We obtain the linearized Schwinger-Keldish effective scalar field
equation at one- and two-loop orders in Sec.~\ref{sec:solve} and
solve it exactly in Sec.~\ref{sec:quantumcorrected}. We calculate
the scalar's quantum corrected power spectrum, its spectral index
and the running of the spectral index at one- and two-loop orders
in Sec.~\ref{sec:Power}. Implications of our results are discussed
in Sec.~\ref{sec:concl}. The Appendices comprise various key steps
of the computations in the paper.

\section{The Model}
\label{sec:model}

In our model, the background metric $g_{\mu\nu}$ describes open
conformal coordinate patch of de~Sitter spacetime. The invariant
line element can be expressed in conformal and comoving
coordinates, respectively, as\beeq ds^2\!=\! g_{\mu\nu} dx^{\mu}
dx^{\nu} \!=\! a^2(\eta) \Bigl[ -d\eta^2 \!+\! d\vec{x} \cdot
d\vec{x} \Bigr]\!=\!-dt^2 \!+\! e^{2 H t} d\vec{x} \cdot d\vec{x}
\; , \label{mtrc}\eneq where the conformal factor \beeq a(\eta) =
-{1 \over H \eta} = e^{H t} \; . \label{cnfrml}\eneq  To exploit
dimensional regularization we work in $D$ spacetime dimensions,
hence the indices $\mu,\nu = 0,1,2, \dots,(D\!-\!1)$. In our
notation: $x^\mu = (x^0,\vec x)$, $x^0\equiv \eta$, and
$\partial_\mu = (\partial_0,\vec\nabla)$.

The state is released in Bunch-Davies vacuum at comoving time
$t\!=\!0$, corresponding to conformal time $\eta
\!=\!\eta_i\!\equiv\!-H^{-1}$. Infinite future
$t\!\rightarrow\!\infty$ corresponds to $\eta\!\rightarrow\!0^-$.
The~conformal factor is normalized to $a\!=\!1$ when the state is
released, hence $a\!>\!1$ throughout the evolution.

We take the Lagrangian density in our model, in terms of
unrenormalized scalar field $\phi$, bare mass $m_0$ and bare
coupling constant $\lambda_0$, as \beeq \mathcal{L}\!=\!
-\frac{1}{2}\sqrt{-g}g^{\mu\nu}
\partial_{\mu} \phi
\partial_{\nu} \phi-\!\frac{m^2_0}2\sqrt{-g}
\phi^2\!-\!\frac{\lambda_0}{4!} \sqrt{-g}\phi^4 \;
.\label{LagDenBARE} \eneq Here, $g$ is the determinant of the
metric $g_{\mu\nu}$. Introducing the renormalized scalar field
\beeq \varphi(x) \equiv \frac{\phi(x)}{\sqrt{Z}} \; ,
\label{fieldstrn}\eneq converts the Lagrangian density
$\mathcal{L}$ to the form \beeq \mathcal{L} \!=\! -\frac{Z}{2}
\sqrt{-g}g^{\mu\nu}\partial_{\mu} \varphi
\partial_{\nu} \varphi-\!\frac{Z m^2_0}{2}\sqrt{-g}
\varphi^2\!-\!\frac{Z^2 \lambda_0}{4!}\sqrt{-g} \varphi^4  \; .
\eneq Renormalization is achieved by decomposing the bare
parameters into renormalized and counter parameters as\beeq Z
\equiv 1\!+\!\delta Z \; , \;\;\; Z m^2_0 \equiv 0 +\!\delta
m^2\;, \;\;\; Z^2 \lambda_0 \equiv \lambda+\!\delta \lambda \;
,\eneq where we enforce the renormalized mass to be zero,
initially. In time, however, quantum processes generate self-mass
\cite{BrOnWo}. So, our scalar field $\varphi(x)$ is classically
massless. Thus, the renormalized Lagrangian density is \beeq
\hspace{-0.6cm}{\cal L}\!=\!-\frac{(1\!+\!\delta
Z)}{2}\sqrt{-g}g^{\mu\nu}
\partial_{\mu} \varphi \partial_{\nu} \varphi\!-\!\frac{\delta
m^2}{2}\sqrt{-g}\varphi^2\!-\!\frac{(\lambda
\!+\!\delta\lambda)}{4!}\sqrt{-g}\varphi^4 \; .
\label{fulLag}\eneq The field strength ($\delta Z$), mass ($\delta
m^2$), and coupling constant ($\delta \lambda$) counterterms
remove divergences at one- and two-loop orders. They are given as
functions of dimensional regularization parameter
$\epsilon\!\equiv\!4\!-\!D$ in Ref.~\cite{BrOnWo}. Varying the
Lagrangian with density (\ref{fulLag}) yields the scalar field
equation \beeq \dd_{\mu} \Bigl(\sqrt{-g} g^{\mu\nu} \dd_{\nu}
\varphi\Bigr)\!=\!\frac{\sqrt{-g}}{1\!+\!\delta Z}\!\left[\delta
m^2\varphi+\!\frac{(\lambda\!+\!\delta \lambda)}{6}\,\varphi^3
\right]\; . \label{phieqn} \eneq Its solution can be given as
\beeq \varphi(x)\!=\!\varphi_0(x)+\!\frac{1}{1\!+\!\delta
Z}\!\int_{\eta_i}^0
\!\!\!d\eta'\!\sqrt{-g(\eta')}\!\int\!\!d^{D-1}x' \,
G(x;x')\!\left[\delta m^2\varphi(x')+\!\frac{(\lambda\!+\!\delta
\lambda)}{6}\,\varphi^3(x') \right] \; , \label{phix} \eneq where
$\varphi_0(x)$ is the solution for the homogeneous equation \beeq
\partial_{\mu} \Bigl(\sqrt{-g} g^{\mu\nu} \partial_{\nu}\varphi_0(x)
\Bigr)\!=\!0\; , \label{phi0}\eneq and the Green's function
$G(x;x')$ is any solution of the equation
\begin{equation}
\partial_{\mu} \Bigl(\sqrt{-g} g^{\mu\nu} \partial_{\nu} G(x;x') \Bigr)
\!=\! \delta^D(x \!-\! x') \; ,
\end{equation}which obeys retarded boundary conditions. It is the free scalar field $\varphi_0(x)$
what mainly concerns us, in this paper. How can we write the most
general $\varphi_0(x)$? In conformal coordinates Eq.~(\ref{phi0})
becomes\beeq\varphi''_0(\eta,\vec{x})+\!(D\!-\!2)\frac{a'}{a}\varphi'_0(\eta,\vec{x})
-\!\nabla^2\varphi_0(\eta,\vec{x})\!=\!0\; ,\label{modedifeq}\eneq
where a prime denotes derivative with respect to the conformal
time $\eta$. To find the solution for Eq.~(\ref{modedifeq}) we
Fourier transform it \beeq
\widetilde{\varphi}_0''(\eta,k)+\!(D\!-\!2)\frac{a'}{a}\widetilde{\varphi}_0'(\eta,k)
+\!k^2\widetilde{\varphi}_0(\eta,k)\!=\!0\; ,\label{fourierphi}
\eneq where
\beeq\widetilde{\varphi}_0(\eta,k)\!=\!\!\int\!d^{D-1}x\,
e^{-i\vec{k}\cdot\vec{x}}\varphi_0(\eta, \vec{x})\; ,\eneq and
look for a function $u(\eta, k)$ which satisfies
Eq.~(\ref{fourierphi}), to wit, \beeq
u''(\eta,k)+\!(D\!-\!2)\frac{a'}{a}u'(\eta,k)
+\!k^2u(\eta,k)\!=\!0\; . \label{modeeqnforu}\eneq Making the
transformation \beeq u(\eta, k)\!\equiv\!a^{-\frac{D-1}2} v(\eta,
k)\; ,\eneq and using Eq.~(\ref{cnfrml}) brings differential
equation (\ref{modeeqnforu}) to the form \beeq v''(\eta,
k)+\!\frac{v'(\eta,
k)}{\eta}+\!\!\left[k^2\!-\!\frac{\nu^2}{\eta^2}\right]\!v(\eta,
k)=0\; ,\label{difeqforv}\eneq where $\nu\!\equiv\!\frac{D-1}{2}$.
Making the change of variable $\eta\!=\!x/k$ reduces
Eq.~(\ref{difeqforv}) to the Bessel's equation of order $\nu$
\beeq\frac{d^2v(x)}{dx^2}+\!\frac{1}{x}\frac{dv(x)}{dx}
+\!\left[1\!-\!\frac{\nu^2}{x^2}\right]\!v(x)=0\; ,\eneq whose
solution is given in terms of the first and second kind Hankel
functions as\beeq v(\eta,k)=\mathcal{A}\mathcal{H}^{(1)}_{\nu}\!
\left(k\eta\right)+\!\mathcal{B}\mathcal{H}^{(2)}_{\nu}\!
\left(k\eta\right)\; .\eneq The coefficients $\mathcal{A}$ and
$\mathcal{B}$ are arbitrary constants. Therefore, the solution
$u(\eta, k)$ of Eq.~(\ref{modeeqnforu}) can be given as \beeq
u(\eta, k)= \mathcal{A} a^{-\frac{D-1}2}
\mathcal{H}^{(1)}_{\frac{D-1}2}\! \left(k\eta\right)+\!\mathcal{B}
a^{-\frac{D-1}2} \mathcal{H}^{(2)}_{\frac{D-1}2}\!
\left(k\eta\right) \; .\label{mode-u-general}\eneq The
Bunch-Davies mode function is associated with the choice of
$\mathcal{B}=0$.

Now, notice the analogy between the kinetic term of our
Lagrangian\beeq
L=\frac{a^{D-2}(\eta)}{2}\!\int\!\!\frac{d^{D-1}k}{(2\pi)^{D-1}}\left[|\widetilde\varphi\,'(\eta,
k)|^2\!-\!k^2|\widetilde\varphi(\eta, k)|^2\right]\; ,\eneq and
the Lagrangian of a simple harmonic oscillator\beeq
L_{SHO}=\frac{m}{2}\!\int\!\! dt \,[\dot{q}^2(t)\!-\!\omega^2
q^2(t)] \;.\eneq The correspondences between
$\widetilde\varphi(\eta, k)\leftrightarrow q(t)$,
$k\leftrightarrow \omega$ and $a^{D-2}\leftrightarrow m$ for each
mode $k$, imply that $\widetilde\varphi(\eta, k)$ behaves like a
simple harmonic oscillator. The well-known solution of the
equation of motion for $q(t)$ in terms of the initial values
$q_I\!\equiv\!q(t_I)$ and $\dot{q}_I\!\equiv\!\dot{q}(t_I)$ is \be
q(t)= q_{\scriptscriptstyle I}\cos(\omega
t)+\!\frac{\dot{q}_{\scriptscriptstyle I}}{\omega}\sin(\omega
t)=\frac{e^{-i\omega t}}{2}\left[q_{\scriptscriptstyle
I}\!+\!\frac{i\dot{q}_{\scriptscriptstyle
I}}{\omega}\right]\!+\!\frac{e^{i\omega
t}}{2}\left[q_{\scriptscriptstyle
I}\!-\!\frac{i\dot{q}_{\scriptscriptstyle I}}{\omega}\right]\;
.\ee The commutator of the operators $[q_{\scriptscriptstyle I},
\dot{q}_{\scriptscriptstyle I}]=i/m$. The annihilation and
creation operators are respectively defined as the operators next
to $e^{-i\omega t}$ and $e^{i\omega t}$,\be
a(\omega)=\sqrt{\frac{m\omega}{2}}\left[q_{\scriptscriptstyle
I}\!+\!\frac{i\dot{q}_{\scriptscriptstyle I}}{\omega}\right]\;\;
,\;\;a^\dagger(\omega)=\sqrt{\frac{m\omega}{2}}\left[q_{\scriptscriptstyle
I}\!-\!\frac{i\dot{q}_{\scriptscriptstyle I}}{\omega}\right]\;
.\ee The normalization factors are chosen to have
$[a,a^\dagger]=1$. Thus, \beeq q(t)\!=\!\frac{e^{-i\omega
t}}{\sqrt{2m\omega}}a(\omega)+\!\frac{e^{i\omega
t}}{\sqrt{2m\omega}}a^\dagger(\omega)\; ,\label{q}\eneq is the
solution of the equation of motion
$\ddot{q}(t)\!+\!\omega^2q(t)=0$ for each $\omega$.

Similarly, by analogy, the solution of equation of motion for the
$\widetilde\varphi(\eta, k)$, for each mode $k$, can be written
as\beeq \widetilde\varphi(\eta, k)\!=\!u(\eta,
k)\alpha(\vec{k})+\!u^*(\eta, k)\alpha^\dagger(-\vec{k})\;
.\label{modesolution}\eneq Here, the mode function $u(\eta, k)$ is
analogous to $\frac{e^{-i\omega t}}{\sqrt{2m\omega}}$, in
Eq.~(\ref{q}), that solves the equation of motion
$\ddot{q}(t)+\omega^2q(t)=0$. Operators $\alpha(\vec{k})$,
$\alpha^\dagger(\vec{k})$ are the annihilation and creation
operators of the scalar and can be expressed in terms of initial
values $a(\eta_I)$, ${u}^*(\eta_I,k)$, ${u'}^*(\eta_I,k)$,
$\widetilde\varphi(\eta_I, \vec{k})$, and
$\widetilde\varphi'(\eta_I,\vec{k})$, analogous to $a$ and
$a^\dagger$. To see this, notice that the canonical commutation
relations \be [\widetilde\varphi(\eta, \vec{k}),
\Pi(\eta,\vec{k}')]\!&=&\![\widetilde\varphi(\eta,
\vec{k}),a^{D-2}(\eta)\widetilde\varphi\,'(\eta,
\vec{k}')]\!=\!i(2\pi)^{D-1}\delta^{D-1}(\vec{k}-\vec{k}')\nonumber\\
\left[ \alpha(\vec{k}),
\alpha^\dagger(\vec{k}')\right]\!\!&=&\!\!(2\pi)^{D-1}\delta^{D-1}(\vec{k}-\vec{k}')\;
, \ee with \beeq\widetilde\varphi\,'(\eta, k)=u'(\eta,
k)\alpha(\vec{k})+\!u'^{*}(\eta, k)\alpha^\dagger(-\vec{k})\;
,\label{modesolutionDERIVAT}\eneq fixes the Wronskian of
differential equation~(\ref{fourierphi}) as\beeq W(\eta,
k)=u(\eta,k){u'}^*(\eta,k)-\!u^{*}(\eta,k)u'(\eta,k)=\frac{i}{a^{D-2}(\eta)}\label{Wronsk}\;
.\eneq Using Eq.~(\ref{Wronsk}), Eqs.~(\ref{modesolution}) and
(\ref{modesolutionDERIVAT}) can be solved for $\alpha(\vec{k})$ as
\beeq
\alpha(\vec{k})=-ia^{D-2}(\eta_I)\!\left[{u'}^*(\eta_I,k)\widetilde\varphi(\eta_I,
\vec{k})-\!{u}^*(\eta_I,k)\widetilde\varphi\,'(\eta_I,
\vec{k})\right]\; . \label{alphak}\eneq The creation operator
$\alpha^\dagger(\vec{k})$ is attained by conjugation.

The coefficient $\mathcal{A}$ in solution (\ref{mode-u-general})
can be obtained by imposing Wronskian (\ref{Wronsk}) to
Eq.~(\ref{mode-u-general}) and using the power series expansion of
the Hankel function $\mathcal{H}^{(1)}_{\nu}$. The result is \beeq
u(\eta,k) \!=\! \sqrt{\frac{\pi}{4H}} a^{-\frac{D-1}2}
\mathcal{H}^{(1)}_\nu\! \Bigl(\frac{k}{H a}\Bigr) \; ,
\label{udef}\eneq where we used the fact that ${H}^{(1)}_\nu$ is
an even function. The Bunch-Davies mode function $u(\eta, k)$ take
a particularly simple form in $D=4$, \beeq u(\eta,k)
\Bigl\vert_{D=4} \!=\! \frac{H}{\sqrt{2 k^3}} \Bigl(1 \!+\! i k
\eta\Bigr) e^{-i k \eta} \!=\! \frac{H}{\sqrt{2 k^3}} \Bigl(1
\!-\! \frac{ik}{H a} \Bigr) \exp\Bigl[\frac{ik}{H a}\Bigr] \; .
\label{u4} \eneq The inverse Fourier transform of mode solution
(\ref{modesolution}) yields the usual free field expansion\beeq
\varphi_0(\eta, \vec{x})=\!\int\!\!\frac{d^{D-1}k}{(2\pi)^{D-1}}
\Bigl\{u(\eta,k) e^{i \vec{k} \cdot \vec{x}} \alpha(\vec{k})\!+\!
u^*(\eta,k) e^{-i\vec{k} \cdot \vec{x}} \a^{\dagger}(\vec{k})
\Bigr\} \; .\label{varphizero} \eneq For $D=4$, we use
Eq.~(\ref{u4}) in Eq.~(\ref{varphizero}) and obtain \beeq
\varphi_0(\eta, \vec{x})=
\!\int\!\!\frac{d^3k}{(2\pi)^3}\frac{H}{\sqrt{2k}}
\Bigl\{e^{-ik\eta+i\vec{k} \cdot
\vec{x}}\left(\!\frac{1}{k}\!+\!i\eta\!\right)\!
\alpha(\vec{k})+\! e^{ik\eta-i\vec{k} \cdot
\vec{x}}\left(\!\frac{1}{k}\!-\!i\eta\!\right)\!
\alpha^\dagger(\vec{k})\Bigr\} \; .\label{varphiD3} \eneq

In this paper, we calculate the one- and two-loop corrections to
the Bunch-Davies mode function~(\ref{u4}) for a MMC scalar with a
quartic self interaction, and then, employ it correct the scalar's
scale-free tree-order power spectrum $\Delta^2_\varphi(t, k)$ at
one- and two-loop orders. As was mentioned earlier, quantum
processes generate effective self-mass which modifies
Eq.~(\ref{phi0}) and, thus, the amplitude $u(\eta, k)$ of the
solution gets quantum corrections. How field equation (\ref{phi0})
is modified at one- and two-loop orders and how $u(\eta, k)$ gets
order-$\lambda$ and -$\lambda^2$ corrections are discussed in the
next two sections.

\section{Linearized Schwinger-Keldish Effective Field Equation}
\label{sec:solve} For a classically massive scalar field
$\phi(x)$, the mass term enters Eq.~(\ref{phi0}) as \beeq
\partial_{\mu} \Bigl(\sqrt{-g} g^{\mu\nu} \partial_{\nu}\phi(x)
\Bigr)\!=\!m^2_0\sqrt{-g}\phi(x)\; .\eneq Exactly the same
procedure for solving Eq.~(\ref{phi0}), outlined in
Sec.~\ref{sec:model}, yields the Bunch-Davies mode function for a
massive scalar. It has the form given in Eq.~(\ref{udef}) with
$\nu\!=\!\sqrt{(\frac{D-1}{2})^2\!-\!\frac{m_0^2}{H^2}}$. Quantum
induced scalar self-mass-squared, however, enters \cite{BrOnWo}
the linearized Schwinger-Keldish effective field equation as a
source term integrated against the scalar, \beeq
\partial_{\mu} \Bigl(\sqrt{-g}
g^{\mu\nu} \partial_{\nu}
\varphi_0(x)\Bigr)\!=\!\int_{\eta_i}^0\!\! d\eta'\!\int\!d^3x'
M^2(x;x')\varphi_0(x')\; ,\label{schkelphizero} \eneq where we
have taken the unregulated limit $D \!=\! 4$ after renormalizing
the scalar self mass-squared. The Schwinger-Keldish formalism
\cite{Sc} is a covariant extension of the Feynman's formulation
which produces expectation values rather than in-out matrix
elements. The reviews in Ref.~\cite{Jor} explain the use of the
formalism.

The self mass squared can be expressed as a series in powers of
the loop counting parameter $\l$, \beeq M^2(x;x')\!=\!
\sum_{\ell=0}^{\infty} \l^\ell\mathcal{M}^2_\ell(x;x')\; .\eneq
The tree-order term in the sum\beeq\mathcal{M}^2_0(x;x')\!=\!0\;
,\label{1e0}\eneq since the scalar is classically massless. We
calculated the one- and two-loop terms $\mathcal{M}^2_1(x;x')$ and
$\mathcal{M}^2_2(x;x')$ in Ref. \cite{BrOnWo}. The one-loop term
is simple\beeq \mathcal{M}^2_1(x;x')\!=\!\frac{H^2}{8\pi^2} \,a^4
\ln(a) \delta^4(x\!-\!x')\; . \label{1el} \eneq The two-loop term
consists of six terms\be
\mathcal{M}^2_2(x;x')\!\equiv\!\sum_{n=1}^6\mathcal{M}^2_{2,n}(x;x')\;
,\label{2el}\ee where\be &&\hspace{-0.5cm}\mathcal{M}^2_{2,
1}(x;x')= \frac{i }{2^{12}\,3\pi^6} a a' \dd^4
\Biggl[\frac{\ln\left(\mu^2 \D x^2_{\scriptscriptstyle
++}\right)}{\D x^2_{\scriptscriptstyle ++}} - \frac{\ln\left(\mu^2
\D x^2_{\scriptscriptstyle
+-}\right)}{\D x^2_{\scriptscriptstyle +-}}\Biggr]\label{m21}\; ,\\
&&\hspace{-0.5cm}\mathcal{M}^2_{2, 2}(x;x')=-\frac{i H^2}{2^9
\pi^6}\! (a a')^2 \dd^2 \Biggl[ \ln\Bigl(\frac{H e^{\frac34}}{2
\mu}\!\Bigr) \Bigg(\frac{\ln\left(\mu^2 \D x^2_{
\scriptscriptstyle ++}\right)}{\D x^2_{\scriptscriptstyle ++}}-
\frac{\ln\left(\mu^2 \D x^2_{ \scriptscriptstyle +-}\right)}{\D
x^2_{\scriptscriptstyle +-}}\Bigg)\Biggr]\label{m22}\; ,\\
&&\hspace{-0.5cm}\mathcal{M}^2_{2, 3}(x;x')=-\frac{i H^2}{2^{11}
\pi^6}\! (a a')^2 \dd^2 \Biggl[\frac{\ln^2\left(\mu^2 \D
x^2_{\scriptscriptstyle ++}\right)}{\D x^2_{ \scriptscriptstyle
++}}-\frac{\ln^2\left(\mu^2 \D x^2_{\scriptscriptstyle
+-}\right)}{\D x^2_{ \scriptscriptstyle +-}}\Biggr]\label{m23}\; ,\\
&&\hspace{-0.5cm}\mathcal{M}^2_{2, 4}(x;x')= -\frac{i H^4}{2^9
\pi^6}\! (a a')^3 \Bigg[\frac{ \ln^2\left(\frac{\sqrt{e}}4 H^2 \D
x^2_{\scriptscriptstyle ++}\right)}{\D x^2_{\scriptscriptstyle
++}}- \frac{ \ln^2\left(\frac{\sqrt{e}}4 H^2 \D
x^2_{\scriptscriptstyle +-}\right)}{\D x^2_{\scriptscriptstyle
+-}}\Bigg]\label{m24}\; ,\\
&&\hspace{-0.5cm}\mathcal{M}^2_{2, 5}(x;x')= \frac{i
H^6}{2^{10}\,3\pi^6}(a a')^4\!\Bigg[ \ln^3\Bigl(\frac{\sqrt{e}}4
H^2 \D x^2_{ \scriptscriptstyle ++}
\Bigr)-\ln^3\Bigl(\frac{\sqrt{e}}4
H^2 \D x^2_{ \scriptscriptstyle +-} \Bigr)\Bigg]\label{m25}\; ,\\
&&\hspace{-0.5cm}\mathcal{M}^2_{2, 6}(x;x')=-\frac{1}{2^9 3\,
\pi^4} \, a^2 \Biggl\{\ln(a) \dd^2 \!-\! \Bigl(2 \ln(a) \!+\!
1\Bigr) H a
\partial_0 \Biggr\} \delta^4(x \!-\! x')\nonumber\\&&\hspace{-0.5cm}
- \frac{H^2}{2^7\,3^2\pi^4}a^4 \Biggl\{4 \ln^3(a) \!+\!
\frac{23}{2} \ln^2(a)\!-\! \Bigl[39 \!+\! 27 \ln\Bigl(\frac{H}{2
\mu} \Bigr) \!-\! 2 \pi^2 \Bigr] \ln(a) \Biggl\} \delta^4(x \!-\!
x')\nonumber\\&&\hspace{-0.5cm}+ \frac{H^2}{2^7 \pi^4}a^4 \Biggl\{
\frac{a^{-3}}{81} \!-\!\sum_{n=1}^\infty
\frac{(n\!+\!5)}{(n\!+\!1)^3}a^{-(n+1)} \!+\! 4\!\sum_{n=1}^\infty
\frac{a^{-(n+2)}}{(n\!+\!2)^3}\!+\! 4\!\sum_{n=1}^\infty\frac{a^{-(n+3)}}{n(n\!+\!3)^3} \Biggr\} \delta^4(x
\!-\! x') \label{m26}\; .\ee In Eqs.~(\ref{m21})-(\ref{m26}) the
coordinate intervals $\Delta x^2_{\scriptscriptstyle ++}$, $\Delta
x^2_{\scriptscriptstyle +-}$, $\Delta x^2_{ \scriptscriptstyle
-+}$ and $\Delta x^2_{\scriptscriptstyle --}$ are defined as
\begin{eqnarray}
\Delta x^2_{\scriptscriptstyle ++}(x;x') \!&\equiv&\! r^2 \!-\!
\Bigl(\vert \eta
\!-\! \eta'\vert \!-\! i \delta \Bigr)^2\; ,\label{plusplus}\\
\Delta x^2_{\scriptscriptstyle +-}(x;x') \!&\equiv&\! r^2 \!-\!
\Bigl(\eta \!-\! \eta' \!+\! i \delta \Bigr)^2\!\!=(\Delta
x^2_{\scriptscriptstyle -+}(x;x'))^* \; , \label{plusminus}\\
\Delta x^2_{\scriptscriptstyle --}(x;x') \!&=&\! (\Delta
x^2_{\scriptscriptstyle ++}(x;x'))^*\; ,
\end{eqnarray} where  $\vec{r}\equiv\vec{x} \!-\! \vec{x}\,'$ and $\delta$ is an infinitesimal positive real
parameter.

The quantum corrected scalar mode function $\Phi(x;\vec{k})$ is
the solution of the linearized effective field
equation~(\ref{schkelphizero}), \beeq \dd_\mu\Bigl(\sqrt{-g}
g^{\mu\nu} \partial_{\nu} \Phi(x;\vec{k})\Bigr) \!=\!\!
\int_{\eta_i}^0 \!\!\!d\eta' \!\!\int\!\!
d^3x'\Bigl\{\sum_{\ell=0}^{\infty} \l^{\ell}
\mathcal{M}^2_{\ell}(x;x') \Bigr\} \Phi(x';\vec{k}) \; .
\label{lineqn} \eneq We solve the effective field equation
perturbatively. Expanding the plane wave solution $\Phi(\eta,k)$
in powers $\lambda$ \beeq \Phi(x;\vec{k})\!\equiv\!\Phi(\eta; k)
e^{i \vec{k} \cdot\vec{x}}\!\equiv\!\sum_{\ell=0}^{\infty}
\l^{\ell} \Phi_{\ell}(\eta,k) e^{i \vec{k} \cdot\vec{x}}\;
,\label{solexp} \eneq and substituting it back into
Eq.~(\ref{lineqn}), we find \beeq a^2 \Bigl[\partial_0^2 \!+\! 2
Ha\partial_0 \!+\! k^2\Bigr] \Phi_{\ell}(\eta,k) \!=\!
-\!\sum_{n=0}^{\ell} \!\int_{\eta_i}^0 \!\!\! d\eta' \!\! \int
\!\! d^3x' \mathcal{M}^2_n(x;x') \Phi_{\ell-n}(\eta',k) e^{i
\vec{k} \cdot (\vec{x}' -\vec{x})} \; .\label{primpert} \eneq Note
that the tree-order ($\ell=0$) solution $\Phi_0(\eta,k)$ is the
Bunch-Davies mode function $u(\eta,k)$ given in Eq.~(\ref{u4}).
Converting Eq.~(\ref{primpert}) for $\Phi_\ell$ from conformal
time $\eta$ to comoving time $t\!=\!-\ln(-H \eta)/H$ yields \beeq
\Bigl[ \frac{\partial^2}{\partial t^2}\!+\! 3 H
\frac{\partial}{\partial t} \!+\! \frac{k^2}{a^2}\Bigr] \Phi_\ell \!=\!
-\frac{1}{a^4}\! \sum_{n=0}^{\ell} \!\int_{\eta_i}^0 \!\!\! d\eta'
\!\! \int \!\! d^3x' \mathcal{M}^2_n(x;x') \Phi_{\ell-n}(\eta',k)
e^{i \vec{k} \cdot (\vec{x}' -\vec{x})} \; . \label{perteq} \eneq
Next, we solve this second order, linear, non-homogeneous
integro-differential equation to get the one- and two-loop quantum
corrections to the Bunch-Davies mode function.

\section{quantum-corrected scalar mode function}
\label{sec:quantumcorrected} The scalar self mass-squared is known
\cite{BrOnWo} at one- and two-loop orders only. Hence, we can
solve Eq.~(\ref{perteq}) for $\Phi_1(\eta, k)$ and $\Phi_2(\eta,
k)$ by applying the Green's function technique.

\subsection{Order $\lambda$ Correction $\Phi_1(\eta, k)$ for the Scalar Mode Function}
\label{sec:Orderlamda}

Using Eqs.~(\ref{1e0}) and (\ref{1el}) in Eq.~(\ref{perteq}), one
obtains the integro-differential equation for the order $\lambda$
correction $\Phi_1(\eta, \vec{k})$, \be \Bigl[
\frac{\partial^2}{\partial t^2} \!+\! 3 H \frac{\partial}{\partial
t} \!+\! \frac{k^2}{a^2}\Bigr] \Phi_1 \!\!&=&\!\!
-\frac{1}{a^4}\!\!\int_{\eta_i}^0\!\!\! d\eta' u(\eta',
k)\!\!\int\!\!d^3x' \mathcal{M}^2_1(x;x')e^{-i \vec{k}
\cdot(\vec{x}-\vec{x}\,')}\nonumber\\
\!\!&=&\!\!-\frac{H^2}{8\pi^2} u(\eta,k)\,\ln(a)\; .
\label{IntegroPhi1}\ee The solution $\Phi_1(\eta,k)$ of
Eq.~(\ref{IntegroPhi1}) can be written as an integral over
comoving time as \beeq
\Phi_1(\eta,k)\!=\!-\frac{H^2}{8\pi^2}\!\!\int_0^t\!\! dt'
G(t,t';k)\, u(\eta',k) \ln\left(a(\eta')\right)\; ,
\label{Phi1}\eneq where the Green's function\beeq
G(t,t';k)=\frac{\theta(t-t')}{W(t',k)}\left[u(\eta, k)u^*(\eta',
k)-u^*(\eta, k)u(\eta', k)\right] \; .\label{Green}\eneq In
Eq.~(\ref{Green}), $\theta(t-t')$ is the Heaviside step function,
and $W(t', k)$ is the Wronskian \beeq W(t', k)\!=\!\dot{u}(\eta',
k)u^*(\eta', k)-u(\eta', k)\dot{u}^*(\eta',
k)\!=\!\frac{-i}{a^3(\eta')}\!=\!iH^3\eta'^3=-ie^{-3Ht'} \;
,\label{Wronskian}\eneq where a dot denotes derivative with
respect to comoving time $t$. Making the change of variable
$t'=\ln(a(\eta'))/H$ and using Eqs.~(\ref{u4}), (\ref{Green}) and
(\ref{Wronskian}) in Eq.~(\ref{Phi1}), we find \be
&&\Phi_1(\eta,k)=-\frac{iH}{8\pi^2}\!\int_1^a\!\!\! da'
a'^2\ln(a')\Bigg[u(\eta, k)|u(\eta', k)|^2\!-\!u^*(\eta, k)u^2(\eta', k)\Bigg]\\
&&=-\frac{iH^3}{16\pi^2k^3}\Bigg\{\!u(\eta, k)\!\!\int_1^a\!\!\!
da' \ln(a')\!\left[a'^2\!+\!\frac{k^2}{H^2}\right]\!\!-\!u^*(\eta,
k)\!\!\int_1^a\!\!\!da'
\ln(a')\!\left[a'\!-\!\frac{ik}{H}\right]^2\!\!\!e^{\frac{2ik}{Ha'}}\!\Bigg\}\;
.\label{Phi1ints}\ee Evaluating the integrals in
Eq.~(\ref{Phi1ints}) gives the one-loop correction to the mode
function\be
&&\hspace{-0.5cm}\Phi_1(\eta,k)\!=\!-\frac{iH^3}{16\pi^2k^3}\Bigg\{\!u(\eta,
k)\!\!\left[a^3\frac{\ln(a)}{3}\!-\!\frac{a^3}{9}\!+\!\frac{1}{9}\!+\!\frac{k^2}{H^2}\Big(a\ln(a)
\!-\!a\!+\!1\Big)\right]\nonumber\\
&&\hspace{-0.5cm}-u^*(\eta,
k)\Bigg[\frac{ik^3}{H^3}\frac{\ln^2(a)}{3} \!+\!\sum_{n \doteq\,
0}^\infty\frac{1}{n!}\left(\frac{2ik}{H}\right)^n\!\!
\Bigg(\!\left[\frac{a^{3-n}}{3\!-\!n}\ln(a)-\frac{(a^{3-n}\!-\!1)}{(3\!-\!n)^2}\right]
\nonumber\\&&\hspace{-0.5cm}-\frac{2ik}{H}\!
\left[\frac{a^{2-n}}{2\!-\!n}\ln(a)-\frac{(a^{2-n}\!-\!1)}{(2\!-\!n)^2}\right]
\!-\!\frac{k^2}{H^2}\!\left[\frac{a^{1-n}}{1\!-\!n}\ln(a)
\!-\!\frac{(a^{1-n}\!-\!1)}{(1\!-\!n)^2}\right]\!\Bigg)\Bigg]\!\Bigg\}\;
, \label{exactPhi1}\; \ee where we use, in the sum, the $\doteq$
symbol to indicate that the terms that diverge for each $n$, as
$n$ runs from $0$ to $\infty$, are to be excluded. In other words,
the terms in the first square brackets for $n\!=\!3$, the terms in
the second square brackets for $n\!=\!2$ and the terms in the
third square brackets for $n\!=\!1$ are to be excluded in the sum.
Equation (\ref{exactPhi1}) is an exact result for
$\Phi_1(\eta,k)$. If we express it in powers of $k/H$, we see that
the lowest order $k$-dependent correction comes in quadratic
order\be\Phi_1(\eta,k)\!\!&=&\!\!\frac{u(0,
k)}{2^4\,3\,\pi^2}\Bigg\{\!\!\!-\!\ln^2(a)\!+\!\frac{2\ln(a)}{3}
\!-\!\frac{2}{9}\!+\!\frac{2a^{-3}}{9}
\!-\!\frac{k^2}{H^2}\Bigg[\frac{1}{5}
\!+\!\frac{\ln^2(a)}{2a^2}\!-\!\frac{7\ln(a)}{3a^2}
\!+\!\frac{16}{9a^2}\!-\!\frac{2}{a^{3}}\!+\!\frac{a^{-5}}{45}\Bigg]\nonumber\\
&&\hspace{-2cm}-\frac{ik^3}{H^3}\Bigg[\frac{2}{27}\!-\!\frac{\ln^2(a)}{3a^3}
\!-\!\frac{2\ln(a)}{9a^3}
\!-\!\frac{2a^{-3}}{27}\Bigg]\!+\!\frac{k^4}{H^4}\Bigg[\frac{3}{140}\!-\!\frac{a^{-2}}{10}
\!+\!\frac{\ln^2(a)}{8a^4}\!-\!
\frac{\ln(a)}{12a^4}\!+\!\frac{5a^{-4}}{18}\!-\!\frac{a^{-5}}{5}
\!+\!\frac{a^{-7}}{1260}\Bigg]\nonumber\\
&&\hspace{10cm}+\mathcal{O}(ik^5/H^5)\!\Bigg\} \;
.\label{power1}\ee Note that, leading logarithm terms in the
$k$-independent part of this result agrees with our putative late
time solution (\ref{oldmodesoln}). The $k$-dependent shifts are
intrinsic to complete solution~(\ref{exactPhi1}). The
$k$-independent shift $-2/9$ in Eq.~(\ref{power1}), which is also
time independent, can be absorbed into a field strength
renormalization~(\ref{fieldstrn}), and hence it is not an
observable. To see this, notice that the full mode
function~(\ref{solexp}) becomes $\bar\Phi(x; k)=\!\bar\Phi(\eta,
k)e^{i\vec{k}\cdot\vec{x}}\!=\![\Phi(\eta;
k)/\sqrt{Z}]e^{i\vec{k}\cdot\vec{x}}$, if we define a new field
$\bar\varphi(x)\!\equiv\!\varphi(x)/\sqrt{Z}$. Expressing
$\sqrt{Z}$ in powers of $\lambda$ as $\sqrt{Z}\!=\!1\!+\!\lambda
Z_1\!+\!\lambda^2 Z_2\!+\!\mathcal{O}(\lambda^3)$, we find \be
&&\hspace{-1.5cm}\bar\Phi(\eta, k)\!=\!\sum_{\ell=0}^{\infty}
\l^{\ell}
\bar\Phi_{\ell}(\eta,k)\!=\!\frac{1}{\sqrt{Z}}\sum_{\ell=0}^{\infty}
\l^{\ell}\Phi_{\ell}(\eta,k)\nonumber\\
&&\hspace{-1cm}=u(\eta, k)\!+\!\lambda\!\left[\Phi_1(\eta,
k)\!-\!Z_1u(\eta, k)\right]\!+\!\lambda^2\!\left[\Phi_2(\eta,
k)\!-\!Z_1\Phi_1(\eta, k)\!-\!Z_2u(\eta,
k)\right]\!+\!\mathcal{O}(\lambda^3)\; .\label{fieldstrength}\ee
The $k$- and $\eta$-independent correction in Eq.~(\ref{power1})
has the form $\lambda u(0,k) Z_1$. Except for the fact that $\eta$
has been taken to $0$, that is exactly the same as a field
strength renormalization. Because $u(\eta,k)$ approaches a nonzero
constant in the late time limit $\eta\rightarrow 0$, and also
because we can unambiguously determine only the late time behavior
of $\Phi_1(\eta, k)$ (the initial state hasn't been perturbatively
corrected), we conclude that, if $Z_1\!=\!\!-1/(2^33^3\pi^2)$, a
$k$-independent constant shift would not be present in the
$\bar\Phi_1(\eta, k)$.

The $k$-dependent constant shifts in Eq.~(\ref{power1}), on the
other hand, cannot be absorbed with a field strength
renormalization and they are, in principle, physical. To fix them
and the terms that redshift like inverse powers of the scale
factor in $\Phi_1(\eta, k)$ unambiguously, however, one must
compute the first order term $\lambda\vert \Omega_1\rangle$ in the
initial state correction \beeq \vert\Delta\Omega\rangle\!\equiv\!
\sum_{n=1}^{\infty}\lambda^n\vert\Omega_n\rangle \;
,\label{initstate} \eneq as in Ref.~\cite{KaOnWo}, and use it to
solve the effective field equation. The terms which grow like
powers of $\ln(a)$ in the one-loop correction $\Phi_1(\eta, k)$
represent the effect of inflationary particle production pushing
the field up its quartic potential and are physical, in principle.
We compute the two-loop correction $\Phi_2(\eta, k)$, in the next
section.

\subsection{Order $\lambda^2$ Correction $\Phi_2(\eta, k)$ for the Scalar Mode Function}
\label{sec:Orderlamdasquarred} Using Eq.~(\ref{perteq}), the
integro-differential equation for $\Phi_2(\eta, k)$ is obtained as
\be &&\hspace{-0.8cm}\Bigl[\frac{\partial^2}{\partial t^2} \!+\! 3
H \frac{\partial}{\partial t} \!+\!
\frac{k^2}{a^2}\Bigr]\Phi_2\!=\!\frac{-1}{a^4}\!\!\int_{\eta_i}^0\!\!\!\!d\eta'\!\!\!\int\!\!
d^3 x'\!\left\{\mathcal{M}_1^2(x;x')\Phi_1(\eta',
k)\!+\!\mathcal{M}_2^2(x;x')u(\eta', k)\right\}\!e^{-i \vec{k}
\cdot(\vec{x}- \vec{x}\,')}\; .\nonumber\\\label{IntegroPhi2}\ee
The right hand side has contributions due to both one- and
two-loop self-mass-squared~terms. The contribution due to one-loop
mass-squared term is trivial to integrate. Using Eqs.~(\ref{1el})
one finds \be
\!-\frac{1}{a^4}\!\!\int_{\eta_i}^0\!\!\!d\eta'\!\!\!\int\!\!
d^3x'\!\mathcal{M}_1^2(x;x')\Phi_1(\eta', k)e^{-i \vec{k}
\cdot(\vec{x}- \vec{x}\,')}\!=\!
-\frac{H^2}{8\pi^2}\ln(a)\Phi_1(\eta, k)\; ,\ee where
$\Phi_1(\eta, k)$ is given in Eqs.~(\ref{exactPhi1}) and
(\ref{power1}). It is useful to break the contribution due to the
two-loop self-mass-squared term into a sum of six integrals: \be
\!-\frac{1}{a^4}\!\!\int_{\eta_i}^0\!\!\!d\eta'\!\!\!\int\!\!
d^3x'\!\mathcal{M}_2^2(x;x')u(\eta', k)e^{-i \vec{k}
\cdot(\vec{x}- \vec{x}\,')}\equiv\!\sum_{n=1}^6
\mathcal{I}_n(\eta,k)\; ,\label{sumofsource}\ee where \beeq
\mathcal{I}_n(\eta,k)\equiv\!-\frac{1}{a^4}\!\!\int_{\eta_i}^0\!\!\!d\eta'\!\!\!\int\!\!
d^3x'\!\mathcal{M}_{2,n}^2(x;x')u(\eta', k)e^{-i \vec{k}
\cdot(\vec{x}- \vec{x}\,')} \; .\label{In}\eneq Recall that
$\mathcal{M}_{2,n}^2(x;x')$ are given in
Eqs.~(\ref{m21})-(\ref{m26}). Thus, Eq.~(\ref{IntegroPhi2}) for
$\Phi_2$ can be written as \beeq \Bigl[\frac{\partial^2}{\partial
t^2} \!+\! 3 H \frac{\partial}{\partial t} \!+\!
\frac{k^2}{a^2}\Bigr]\Phi_2\!=\!\mathcal{S}(\eta, k)\;
,\label{diffeqmainPHI2}\eneq where we define the non-homogeneous
(source) term of the differential equation as \beeq
\mathcal{S}(\eta,
k)\equiv-\frac{H^2}{8\pi^2}\ln(a(\eta))\Phi_1(\eta,
k)\!+\!\sum_{n=1}^6 \mathcal{I}_n(\eta,k) \;
.\label{STermIntegroDE}\eneq We explicitly evaluate the first
integral in the source term, \beeq \mathcal{I}_1(\eta,
k)=\frac{-i}{2^{12} \, 3 \, \pi^6 \, a^3}\!\! \int\!\! d^4x' a'
\left\{\!\partial^4\!\left[\frac{\ln(\mu^2 \Delta x^2_{
\scriptscriptstyle ++})}{\Delta x^2_{\scriptscriptstyle ++}} -
\frac{ \ln(\mu^2 \Delta x^2_{\scriptscriptstyle +-})}{\Delta
x^2_{\scriptscriptstyle +-}}\right]\!\right\} u(\eta',k) e^{-i
\vec{k} \cdot(\vec{x}- \vec{x}\,')} \; ,\label{I1first} \eneq to
illustrate the relevant calculation techniques in this section,
and outline the key steps for the remaining five in the
Appendices. The first step is to extract another derivative making
use of the identity\beeq \frac{\ln(\mu^2\Delta x^2)}{\Delta
x^2}=\frac{\partial^2}{8}\left[\ln^2(\mu^2\Delta
x^2)-2\ln(\mu^2\Delta x^2)\right]\; .\label{logIdder}\eneq Then,
one exploits the cancelation between the $++$ and $+-$ terms
except within the past light-cone, using \beeq \ln\!\left[\mu^2\D
x^2_{\scriptscriptstyle +\pm}\right]
\!=\!\ln\!\left[\mu^2(\D\eta^2 \!-\! r^2)\right]\!\pm\! i\pi
\theta(\D \eta^2 \!-\! r^2) \; . \label{id4} \eneq The result is
\be\hspace{-1cm}\mathcal{I}_1(\eta, k) \!&=&\! \frac{-ie^{-i
\vec{k} \cdot \vec{x}}}{2^{15} \, 3\, \pi^6\,a^3}\!\!
\int\!\!d^4x' a' \left\{\!\partial^6\!\left[\matrix{ \ln^2(\mu^2
\Delta x^2_{\scriptscriptstyle ++}) \!-\! 2 \ln(\mu^2 \Delta x^2_{
\scriptscriptstyle ++}) \cr - \ln^2(\mu^2 \Delta
x^2_{\scriptscriptstyle +-}) \!+\! 2 \ln(\mu^2 \Delta
x^2_{\scriptscriptstyle +-})}\right]\!\right\} u(\eta',k) e^{i
\vec{k}
\cdot\vec{x}\,'} \\
\hspace{-1cm}&=&\!\frac{e^{-i\vec{k}\cdot\vec{x}}}{2^{13} \, 3\,
\pi^5}\frac{\partial^6}{a^3}\!\!\int\!\!d\eta'
a'u(\eta',k)\!\!\int\!\! d^3x' e^{i \vec{k}
\cdot\vec{x}\,'}\theta(\D \eta \!-\! r)\!\left\{\ln\Bigl[ \mu^2
(\Delta \eta^2 \!-\! r^2) \Bigr] \!-\! 1\right\} \; . \ee The next
step is to perform the angular integrations using
\begin{equation}
\int\!\!d^3x' e^{i \vec{k} \cdot \vec{x}\,'} \!f(\Delta x) = 4 \pi
e^{i \vec{k} \cdot \vec{x}} \!\int_0^{\infty}\!\!dr r^2
\frac{\sin(k r)}{k r} f(r) \label{angular}\; .
\end{equation}
We can also pass the spatial phase factor through the derivative,
\beeq
\partial^6 e^{i \vec{k} \cdot \vec{x}} \!=\! - e^{i \vec{k} \cdot \vec{x}}
\Bigl( \partial_0^2 \!+\! k^2 \Bigr)^3 \; , \eneq to reach the
form, \beeq \mathcal{I}_1(\eta, k) \!=\! \frac{-1}{2^{11} \, 3\,
\pi^4} \, \frac{\left(\partial_0^2 \!+\! k^2\right)^3}{a^3 k}\!
\int_{\eta_i}^\eta \!\!d\eta' a' u(\eta',k)\!\!
\int_0^{\D\eta}\!\! dr r \sin(k r) \Bigl\{ \ln\Bigl[ \mu^2(
{\Delta \eta}^2-\! r^2) \Bigr] \!-\! 1 \Bigr\} \; .\label{I1}
\eneq The radial integration in Eq.~(\ref{I1}) involves a
combination of special functions. First, we make the change of
variable, $r \equiv {\Delta \eta} z\equiv \frac{\alpha}{k}z$, and
obtain \be &&\hspace{-2.5cm}\int_0^{\Delta \eta}\!\! dr r \sin(k
r) \Bigl\{ \ln\Bigl[ \mu^2 ({\Delta \eta}^2 \!-\! r^2) \Bigr]
\!-\! 1 \Bigr\}\nonumber\\\!&=&\!{\Delta \eta}^2\!\!\int_0^1\!\!
dz z \sin(k {\Delta \eta} z) \Bigl\{ 2 \ln(\mu
{\Delta \eta}) \!-\! 1 \!+\! \ln(1 \!-\! z^2) \Bigr\} \\
\!&=&\!\frac1{k^2} \Bigl[ \sin(k {\Delta \eta}) \!-\! k {\Delta
\eta} \cos(k {\Delta \eta}) \Bigr]\!\Bigl[ 2 \ln(\mu {\Delta
\eta}) \!-\! 1 \Bigr] \!+\! {\Delta \eta}^2 \xi(k {\Delta \eta})
\; ,\label{rad} \ee where we define\begin{eqnarray} \xi(k {\Delta
\eta}) \!&=&\!\xi(\alpha)\equiv\!\int_0^1 \!\!dz z \sin(\alpha z)
\ln(1 \!-\! z^2) \; ,\label{xialphafirst} \\
&=&\!\!\frac2{\alpha^2} \sin(\alpha) \!-\! \frac1{\alpha^2}
[\cos(\alpha) \!+\! \alpha \sin(\alpha)]\!\left[ {\rm si}(2
\alpha) \!+\! \frac{\pi}2 \right]
\nonumber \\
&&\hspace{1.6cm} + \frac1{\alpha^2} [\sin(\alpha) \!-\! \alpha
\cos(\alpha)]\!\left[ {\rm ci}(2 \alpha) \!-\!
\ln\Bigl(\frac{\alpha}2\Bigr) \!-\! \gamma \right] \;
.\label{xialphasecond}
\end{eqnarray}
See Appendix \ref{App:xi} for the evaluation of $\xi(\alpha)$. The
sine and cosine integrals are defined in Eqs. (\ref{siint}) and
(\ref{ciint}), respectively. Inserting Eq.~(\ref{rad}) into
Eq.~(\ref{I1}) and applying the derivatives using \be
\partial_0\!\!\int_{\eta_i}^\eta\!\! d\eta' f(\eta,\eta\,')
\!=\!f(\eta,\eta)\!+\!\!\int_{\eta_i}^\eta \!\!d\eta'
\frac{\partial f(\eta,\eta\,')}{\partial\eta}\; ,
\label{derivativeONintegral}\ee we obtain \be
&&\hspace{-1cm}\mathcal{I}_1(\eta, k) \!=\!\frac{-1}{2^{9} \, 3 \,
\pi^4}\,\frac{k^2}{a^3}
\Bigg\{\!a\,u(\eta,k)\!+\!k\!\!\int_{\eta_i}^\eta\!\!\!d\eta' a'
u(\eta',k)\!\left[\frac{\cos(\a)}{\a}\!-\!\frac{\sin(\a)}{\a^2}
\!+\!\ln\left(2\mu\Delta\eta\right)\sin(\a)\right]\nonumber\\&&\hspace{5cm}+\!\left[2\partial_0
\!+\!\frac{\partial_0^3}{k^2}\right]\!\int_{\eta_i}^\eta\!\!\!d\eta'
a' u(\eta',k)\ln\left(2\mu\Delta\eta\right)\cos(\alpha) \Bigg\}\;
. \ee Expanding the logarithm,
\beeq\ln\!\left(2\mu\Delta\eta\right)\!=\!\ln\!\Big(\!\frac{2\,\mu}{Ha'}\!\Big)
\!+\!\ln\!\Big(\!1\!-\!\frac{a'}{a}\!\Big)\!=\!\ln\!\Big(\!\frac{2\,\mu}{Ha'}\!\Big)
\!-\!\sum_{n=1}^\infty\frac{{a'}^n}{na^n}\; ,
\label{logexpansion}\eneq evaluating the integrals, and taking the
derivatives using
$\partial_0\!=\!\partial_\eta\!=\!Ha^2\partial_a$ one can express
$\mathcal{I}_1(\eta, k)$ in powers of $k/a$ as \be
&&\hspace{-0.8cm}\mathcal{I}_1(\eta, k)\!=\!\frac{-1}{2^{9} \, 3
\, \pi^4}\frac{H}{\sqrt{2k^3}}\!
\Bigg\{\!H^2\!\left[2\ln\!\Big(\!\frac{2\,\mu}{Ha}\!\Big)
\!-\!3\!-\!\sum_{n=2}^\infty\frac{(n\!-\!1)n\,a^{-(n+1)}}{n\!+\!1}\right]
\!+\!\frac{k^2}{a^2}\!\Bigg[\ln\!\Big(\!\frac{2\,\mu}{Ha}\!\Big)
\!-\!\frac{1}{2}\nonumber\\
&&\hspace{0.6cm}-\frac{1}{2}\!\sum_{n=0}^\infty\frac{(n^2\!+\!3n\!+\!4)\,a^{-(n+1)}}{n\!+\!1}\Bigg]
\!\!+\!\frac{ik^3}{Ha^3}\!\left[\frac{2}{3}\ln\!\Big(\!\frac{2\,\mu}{H}\!\Big)
\!-\!1\!-\!\frac{1}{3}\!\sum_{n=0}^\infty\frac{(n\!+\!2)(n\!+\!3)a^{-(n+1)}}{n\!+\!1}\right]\nonumber\\
&&\hspace{0.6cm}-\frac{k^4}{H^2a^4}\!\left[\frac{1}{4}\ln\!\Big(\!\frac{2\,\mu}{Ha}\!\Big)
\!-\!\frac{7}{8}\!-\!\frac{1}{8}\!\sum_{n=0}^\infty\frac{(n^2\!+\!7n\!+\!8)a^{-(n+1)}}{n\!+\!1}\right]
\!+\!\mathcal{O}\Big(\frac{ik^5}{H^3}\Big)\!\Bigg\}\;
.\label{I1exp}\ee Calculations of $\mathcal{I}_n(\eta, k)$ for
$2\leq n\leq 6$ are given in Appendix \ref{App:In}. Using Eqs.
(\ref{power1}), (\ref{I1exp}), (\ref{I2exp}), (\ref{I3exp}),
(\ref{I4exp}), (\ref{I5exp}), (\ref{I6exp}), one obtains an exact
expression for the source term $\mathcal{S}(\eta, k)$ of
Eq.~(\ref{diffeqmainPHI2}). The result is given in
Appendix~\ref{App:S}.

The solution $\Phi_2(\eta,k)$ of Eq.~(\ref{diffeqmainPHI2}) can be
written as an integral over comoving time \beeq
\Phi_2(\eta,k)=\!\!\int_0^t\!\!\!dt'\, G(t,t';k)\,S(\eta',k)\;
.\label{Phi2intso}\eneq  Inserting the Green's function
$G(t,t';k)$ given in Eq.~(\ref{Green}) into Eq.~(\ref{Phi2intso})
yields \be &&\hspace{-1cm}\Phi_2(\eta,k)\!=\!\frac{i}{\sqrt{2k^3}}
\Bigg\{\!u(\eta, k)\!\!\int_1^a\!\!\!da\,'\Bigg[
a'^2\!+\!\frac{ika'}{H} \Bigg]\! \exp\!\Big(\!\frac{-i\,k}{H
a'}\!\Big)S(\eta',k)\nonumber\\
&&\hspace{4cm}-u^*(\eta,k)\!\!\int_1^a\!\!\!da\,'\Bigg[a'^2\!-\!\frac{ika'}{H}\Bigg]\!
\exp\!\Big(\!\frac{i\,k}{H a'}\!\Big)S(\eta',k)\!\Bigg\}\; .
\label{exactPhi2}\ee Using Eq.~(\ref{inhomogen}) in
Eq.~(\ref{exactPhi2}) one finds an exact expression for the
two-loop correction $\Phi_2(\eta, k)$. If we express it in powers
of $k/H$, we see that the lowest order $k$-dependent correction
comes in quadratic order\be
&&\hspace{-0.5cm}\Phi_2(\eta,k)\!=\!\frac{u(0,k)}{2^8\,3^3\,\pi^4}
\Bigg\{\!\frac{11\ln^4(a)}{2}-\ln^3(a)
\!+\!\left(27\ln\!\Big(\frac{2\mu}{H}\Big)\!-\!\frac{547}{12}\!+\!4\pi^2
\!+\!\frac{2}{3a^3}\right)\ln^2(a)\nonumber\\
&&\hspace{-0.5cm}+\!\left(\!\!18\ln^2\!\Big(\frac{2\mu}{H}\!\Big)
\!-48\ln\!\Big(\frac{2\mu}{H}\!\Big)\!+\!\frac{817}{18}
\!+\!\frac{5\pi^2}{3}
\!-\!48\zeta(3)\!+\!\sum_{p=2}^\infty\!\sum_{n=1}^{p-1}\!\frac{72}{n(p-n)p(p+3)}
\!+\!\frac{\frac{43}{18}\!+\!\frac{4\pi^2}{3}}{a^{3}}\right)\!\ln(a)\nonumber\\
&&\hspace{-0.5cm}+\,\mathcal{C}_0\!+\!\mathcal{O}(a^{-2})
\!+\!\frac{k^2}{H^2}\!\Bigg[\!\Bigg(\frac{3}{5}
\!+\!\frac{11\ln^2(a)}{4a^2}\!-\!\frac{45\ln(a)}{2a^2}
\!+\!\frac{\frac{27}{2}\ln\!\left(\frac{2\mu}{H}\!\right)
\!+\!\frac{1181}{24}\!+\!2\pi^2}{a^2}+\!\frac{6}{a^3}
\!-\!\frac{a^{-5}}{15}\!\Bigg)\!\ln^2(a)\nonumber\\
&&\hspace{-0.5cm}+\Bigg(\!\frac{12983}{750}\!-\!\frac{4\pi^2}{5}
\!+\!\frac{9\ln^2\!\left(\frac{2\mu}{H}\!\right)
\!-\!78\ln\!\left(\frac{2\mu}{H}\!\right)\!-\!\frac{2705}{36}
\!-\!\frac{43\pi^2}{6}\!-\!24\zeta(3)}{a^2}
\!+\!\sum_{p=2}^\infty\sum_{n=1}^{p-1}\!\frac{36\,a^{-2}}{n(p-n)p(p+3)}\nonumber\\
&&\hspace{-0.5cm}-\,\frac{\frac{151}{2}-4\pi^2}{a^3}-\frac{\frac{263}{900}
+\frac{2\pi^2}{15}}{a^5}\Bigg)\!\ln(a)\!+\!\mathcal{C}_2\!+\!\mathcal{O}(a^{-2})
\Bigg]\!+\!\frac{ik^3}{H^3}\Bigg[\Big(\frac{2}{9}\!+\!\frac{\ln^2(a)}{2a^3}
\!+\!\frac{13\ln(a)}{9a^3}\nonumber\\
&&\hspace{-0.5cm}-\,\frac{9\ln\!\left(\frac{2\mu}{H}\!\right)+\frac{1921}{252}}{a^3}\Big)\!\ln^2(a)
\!+\!\Big(\frac{457}{54}\!-\!\frac{4\pi^2}{9}\!-\!\frac{6\ln^2\!\!\left(\frac{2\mu}{H}\!\right)
\!-\!4\!\ln\!\left(\frac{2\mu}{H}\!\right)\!-\!\frac{5051}{189}
\!-\!\frac{13\pi^2}{3}\!+\!\!16\zeta(3)}{a^3}\nonumber\\
&&\hspace{-0.5cm}+\!\sum_{p=4}^\infty\!\sum_{n=1}^{p-1}\!\frac{12\,(p\!+\!4)\,a^{-3}}{n(p\!-\!n)(p\!-\!3)(p\!-\!2)p}\Big)\!\ln(a)
\!+\!\mathcal{C}_3\!+\!\mathcal{O}(a^{-2})\!\Bigg]\!-\!\frac{k^4}{H^4}\Bigg[\Big(\frac{9}{140}
\!-\!\frac{3}{10a^2}\!+\!\frac{11\ln^2(a)}{16a^4}
\!-\!\frac{\!\ln(a)}{8a^4}\nonumber\\
&&\hspace{-0.5cm}+\,\frac{\frac{27}{8}\ln\!\left(\frac{2\mu}{H}\!\right)
\!+\!\frac{1109}{96}\!+\!\frac{\pi^2}{2}}{a^4}\!+\!\frac{3a^{-5}}{5}\!-\!\frac{a^{-7}}{420}\Big)\!\ln^2(a)\!+\!
\Big(\frac{540817}{171500}\!-\!\frac{6\pi^2}{35}-\frac{\frac{11183}{1500}-\frac{2\pi^2}{5}}{a^2}
\!+\!\frac{6}{a^3}\nonumber\\
&&\hspace{-0.5cm}+\,\frac{\frac{9}{4}\ln^2\!\left(\frac{2\mu}{H}\!\right)
\!-\!6\ln\!\left(\frac{2\mu}{H}\right)\!+\!\frac{1033}{144}
\!+\!\frac{5\pi^2}{24}\!-\!6\zeta(3)}{a^4}\!+\!\sum_{p=2}^\infty\!\sum_{n=1}^{p-1}\!\frac{9a^{-4}}{n(p-n)p(p+3)}
\!-\!\frac{\frac{707}{100}\!-\!\frac{2\pi^2}{5}}{a^5}\nonumber\\
&&\hspace{-0.5cm}-\,\frac{\frac{2081}{176400}\!+\!\frac{\pi^2}{210}}{a^7}
\Big)\!\ln(a)\!+\!\mathcal{C}_4\!+\!\mathcal{O}(a^{-2})\Bigg]
\!+\!\mathcal{O}\Big(\frac{ik^5}{H^5}\Big)\!\Bigg\}\;
,\label{Phi2}\ee where $\mathcal{C}_0$, $\mathcal{C}_2$,
$\mathcal{C}_3$ and $\mathcal{C}_4$ are time independent constant
terms. The quartic and cubic order logarithm terms in the
$k$-independent part of this result agrees with our putative late
time solution (\ref{oldmodesoln}). The $k$-independent shift
$\mathcal{C}_0$, which is also time independent, can be absorbed
into a field strength renormalization and is not an observable.
(If $Z_2\!=\!(27\mathcal{C}_0\!-\!4)/(2^8 3^6 \pi^4)$ in
Eq.~(\ref{fieldstrength}), a $k$-independent constant shift would
not be present in the $\bar\Phi_2$.) To fix the $k$-dependent
constant shifts $\mathcal{C}_2$, $\mathcal{C}_3$, $\mathcal{C}_4$
and the terms that redshift like inverse powers of the scale
factor unambiguously, one needs to compute the second order term
$\lambda^2 \Bigl\vert \Omega_2\Bigr\rangle$ in initial state
correction~(\ref{initstate}), as in Ref.~\cite{KaOnWo}, and use it
to solve the effective field equation. The terms in powers of
infrared logarithms are physical, in principle.

Thus, the one- and two-loop corrected scalar mode function can now
be expressed as\beeq \Phi(x, k)\!=\!\Phi(\eta,
k)e^{i\vec{k}\cdot\vec{x}}\!=\!\Big(u(\eta,
k)\!+\!\lambda\Phi_1(\eta, k)\!+\!\lambda^2\Phi_2(\eta,
k)\!+\mathcal{O}(\lambda^3)\Big)e^{i\vec{k}\cdot\vec{x}}\;
,\label{Phi2loop}\eneq where $u(\eta, k)$, $\Phi_1(\eta, k)$ and
$\Phi_2(\eta, k)$ are obtained in Eqs.~(\ref{u4}),
(\ref{exactPhi1})-(\ref{power1}) and (\ref{Phi2}), respectively.
In general, $k$-dependent physical shifts in a mode function tilt
the power spectrum. {\it Assuming} we can measure the tree-order
power spectrum of our scalar $\varphi(x)$, we expect that the
$k$-dependent shifts in the loop corrections $\Phi_\ell(\eta, k)$
ought to induce a tilt, which may be observed, in the power
spectrum. In Sec.~\ref{sec:Power}, we use the one- and two-loop
mode function corrections, $\Phi_1(\eta, k)$ and $\Phi_2(\eta,
k)$, obtained in Eqs.~(\ref{power1}) and (\ref{Phi2}), to compute
the scalar's power spectrum at one- and two-loop orders.

\section{quantum-corrected  power spectrum}
\label{sec:Power} The power spectrum $\mathcal{P}(t, k)$ is a
fundamental theoretical quantity in cosmology which measures
typical amplitude of fluctuations in a field as a function of
comoving wave number~$k$. It is defined as the Fourier transform
of the equal-time two-point correlation function of~the field, \be
\mathcal{P}_\varphi(t, k)\!\equiv\!\!\int\!\! d^3x\,
e^{-i\vec{k}\cdot\vec{x}}\langle\Omega|\varphi(t, \vec{x})
\varphi(t, 0)|\Omega\rangle\; . \label{varianceP}\ee The excess
power in a bin of size $dk$ centered at $k$ is associated with
$\mathcal{P}_\varphi(t, k) d^3k/(2\pi)^3$. After integrating it
over all orientations of $\vec{k}$, an alternative measure of
power $\Delta^2_\varphi (t, k)$ in a mode $k$ is obtained\beeq
\frac{k^2dk}{2\pi^2}\mathcal{P}_\varphi(t, k) \!\equiv\!
\frac{dk}{k}\Delta^2_\varphi (t, k) \; .\label{Deltaphi}\eneq
Employing the free-field expansion~(\ref{varphizero}) in
Eqs.~(\ref{varianceP}) and (\ref{Deltaphi}) yields the tree-order
power spectrum as \beeq\Delta^2_\varphi(t,
k)\!=\!\frac{k^3}{2\pi^2}\mathcal{P}_\varphi(t,
k)\!=\!\frac{k^3}{2\pi^2}|u(t, k)|^2\; ,\label{PowerDEfn}\eneq
where $u(t, k)$ is the tree-order mode function given in
Eq.~(\ref{u4}). Hence, \beeq \Delta^2_\varphi(t,
k)\!=\!\frac{H^2}{4\pi^2}\!\left(\!1\!+\!\frac{k^2}{H^2}a^{-2}\right)\;
.\label{PowerDELTA}\eneq This yields the well known
scale-invariant power spectrum ${H^2}/{4\pi^2}$, in the late time
limit.

A non-linear generalization of $\Delta^2_\varphi(t, k)$ is
realized by replacing the tree-order Bunch-Davies mode function
$u(t, k)$ in free field expansion~(\ref{varphizero}) with the
quantum corrected mode function $\Phi(t, k)$ defined in
Eq.~(\ref{solexp}). Employing this loop corrected field expansion
in Eqs.~(\ref{varianceP}) and (\ref{Deltaphi}) yields the quantum
corrected power spectrum \cite{MiPa} as\beeq \Delta^2_\varphi(t,
k)\!=\!\frac{k^3}{2\pi^2} |\Phi(t, k)|^2\;
,\label{quantcorrmode}\eneq where the amplitude squared for each
mode is\be \hspace{-1.5cm}|\Phi(t, k)|^2 \!=\!|u(t,
k)|^2\!\!&+&\!\!\lambda \Big[u(t, k)\Phi_1^*(t, k)\!+\!u^*(t,
k)\Phi_1(t, k)\Big]\nonumber\\
\!\!&+&\!\!\lambda^2 \Big[|\Phi_1(t, k)|^2\!+\!u(t, k)\Phi_2^*(t,
k)\!+\!u^*(t, k)\Phi_2(t, k)\Big]\!+\!\mathcal{O}(\lambda^3) \;
.\label{Phisquared}\ee Precisely measurable cosmological
observables may be sensitive to signals originated from tiny
quantum effects. Motivated by the hope to resolve such signals
theorists intensively studied
---potentially enhanced--- loop corrections to inflationary
correlators in the recent past \cite{LoopRefs}. In this section,
we calculate the one- and two-loop corrections to the power
spectrum of the MMC scalar $\varphi(x)$ with quartic
self-interaction during $\Lambda$-driven de Sitter inflation.
Inserting Eq.~(\ref{Phisquared}) into Eq.~(\ref{quantcorrmode}),
and using Eqs.~(\ref{u4}), (\ref{power1}) and (\ref{Phi2}) after
field strength renormalization, yields \be
&&\hspace{-0.4cm}\Delta^2_\varphi(t,
k)\!=\!\frac{H^2}{4\pi^2}\Bigg\{\!1\!+\!\frac{k^2}{H^2}a^{-2}\!-\!\frac{\lambda}{2^3\,3\,\pi^2}\Bigg[\!\ln^2(a)
\!-\!\frac{2\ln(a)}{3}\!-\!\frac{2}{9a^{3}}\nonumber\\
&&\hspace{-0.4cm}+\frac{k^2}{H^2}\Big(\frac{1}{5}\!+\!\frac{\ln^2(a)}{a^2}
\!-\!\frac{8\!\ln(a)}{3a^2}\!+\!\frac{16}{9a^2}\!-\!\frac{2}{a^3}\!-\!\frac{4a^{-5}}{45}\!\Big)
\!-\!\frac{k^4}{H^4}\Big(\!\frac{3}{140}\!-\!\frac{a^{-2}}{5}\!+\!\frac{\ln(a)}{a^4}
\!-\!\frac{11a^{-4}}{18}\!+\!\frac{4a^{-5}}{5}\!-\!\frac{4a^{-7}}{105}\!\Big)\nonumber\\
&&\hspace{-0.4cm}+\mathcal{O}\Big(\!\frac{k^6}{H^6}\!\Big)\!\Bigg]
\!\!+\!\frac{\lambda^2}{2^7\,3^2\,\pi^4}\!\Bigg[\frac{7\!\ln^4(a)}{3}
\!-\!\ln^3(a)\!-\!\Big[\frac{539}{36}\!-\!\frac{4\pi^2}{3}
\!-\!9\ln\!\Big(\!\frac{2\mu}{H}\Big)\!\Big]\!\ln^2(a)\!+\!\!\Big[\frac{817}{54}
\!+\!\frac{5\pi^2}{9}\!+\!6\ln^2\!\Big(\!\frac{2\mu}{H}\Big)\nonumber\\
&&\hspace{-0.4cm}-16\ln\!\Big(\!\frac{2\mu}{H}\Big)\!-\!16\zeta(3)
\!+\!\!\sum_{p=2}^\infty\!\sum_{n=1}^{p-1}\frac{24}{n(p\!-\!n) p\,(p\!+\!3)}\!+\!\Big(\!\frac{17}{18}
\!+\!\frac{4\pi^2}{9}\!\Big)a^{-3}\Big]\!\ln(a)\!-\!\frac{27a^{-2}}{2}
\!+\!\mathcal{O}(a^{-3})
\nonumber\\
&&\hspace{-0.4cm}+\frac{k^2}{H^2}\Bigg(\!\frac{2\ln^2(a)}{5}
\!+\!\Big[\frac{12683}{2250}\!-\!\frac{4\pi^2}{15}\Big]\!\ln(a)\!+\!\Big[\frac{7\ln^2(a)}{3}
\!-\!\frac{31\!\ln(a)}{3}\!+\!9\ln\!\Big(\!\frac{2\mu}{H}\Big)\!+\!\frac{437}{36}
\!+\!\frac{4\pi^2}{3}\Big]\!\frac{\ln^2(a)}{a^2}\nonumber\\
&&\hspace{-0.4cm}+\Big[6\ln^2\!\Big(\!\frac{2\mu}{H}\Big)\!-\!34\!\ln\!\Big(\!\frac{2\mu}{H}\Big)
\!\!-\!16\zeta(3)\!-\!\frac{56}{3}
\!-\!\frac{19\pi^2}{9}\!+\!\!\sum_{p=2}^\infty\!\sum_{n=1}^{p-1}\!\frac{24}{n(p\!-\!n)p\,(p\!+\!3)}
\!+\!\mathcal{O}(a^{-1})\Big]\!\frac{\ln(a)}{a^2}\!+\!\frac{\mathcal{C}_2}{3}\nonumber\\
&&\hspace{-0.4cm}+\mathcal{O}(a^{-2})\!\!\Bigg)
\!-\!\frac{k^4}{H^4}\Bigg(\!\frac{3\ln^2(a)}{70}
+\!\Big[\frac{533467}{514500}\!-\!\frac{2\pi^2}{35}\Big]\!\ln(a)
\!-\!\Big[\frac{2}{5}\!+\!\mathcal{O}\Big(\frac{\ln(a)}{a^2}\!\Big)\Big]\frac{\ln^2(a)}{a^2}\nonumber\\
&&\hspace{3.5cm}+\Big[\frac{4\pi^2}{15}\!-\!\frac{10883}{2250}\!+\!\mathcal{O}(a^{-1})\Big]\frac{\ln(a)}{a^2}
\!-\!\frac{1}{50}\!+\!\frac{\mathcal{C}_4}{3}\!+\!\mathcal{O}(a^{-2})\!\!\Bigg)
\!+\!\mathcal{O}\Big(\!\frac{k^6}{H^6}\!\Big)\Bigg]\!\Bigg\}\;
.\label{powerspectrum}\ee Thus, the quantum corrections induce
$k$-dependence to tree-order power spectrum~(\ref{PowerDELTA}) in
even powers of $k/H$, in both loop orders. To fix the
$k$-dependent constant terms in the spectrum unambiguously, the
order-$\lambda$ and -$\lambda^2$ initial state corrections must be
worked out in the mode function stage of the calculation, as
discussed in Secs.~\ref{sec:Orderlamda}-B. We read off from
Eq.~(\ref{powerspectrum}) that, at one-loop order, the leading
term is quadratic in time, whereas the $k$-dependent terms are
constants. At two-loop order, on the other hand, the leading term
is quartic in time, whereas the $k$-dependent terms, in leading
order, are quadratic in time.

The spectral index $n$ measures the variation of the power
spectrum of fluctuations in a field with scale. We may quantify
the spectral index $n_\varphi$ of our scalar field $\varphi(x)$
the same way as for the graviton, \beeq n_\varphi(t,
k)\!=\!\frac{d\ln\left(\Delta^2_\varphi(t, k)\right)}{d\ln(k)} \;
.\label{spectr}\eneq The prediction we would like to make for a
cosmological observable like the spectral index $n$ of
fluctuations in a field is for a realistic evolution in which
inflation ends---unlike the pure de Sitter background of our model
where inflation lasts forever. If we had the exact scenario we
could compute whatever a full mode function imprints itself upon
an observable at the time of recombination, a long
---but finite--- time after the end of inflation. Unfortunately,
we can't get what we really want for our spectator scalar
$\varphi(x)$ for many reasons:

(i) We don't know how the scalar
---which is not measured today--- communicates with observables like
the metric fluctuation, which is measured;

(ii) We don't know the actual expansion history in which the
scalar evolves;

(iii) We don't have even the tree-order solution for our mode
function in an arbitrary expansion history, etc. One common
technique to bypass such unknown details is the WKB method. One
treats the {\it tree-order} mode function as evolving in an
arbitrary background in this method, and finds that it approaches
a constant for each mode after horizon crossing. (The fluctuation
is frozen to a constant value outside horizon.) That constant can
be expressed, in terms of the value of the mode function at time
$t\!=\!t_k$ of first horizon crossing of the fluctuation with mode
$k$. This is an approximate way of getting the constant that the
mode function, and hence an observable defined via the mode
function, approaches at tree-order.

Now, let's consider our quantum loop corrections some of which are
time dependent due to the infrared logarithms. They really do not
approach constants, although they only grow very slowly. If the
spacetime background were pure de Sitter, they would continue to
grow forever. The background, however, cannot really be pure de
Sitter, because inflation had to end. To estimate how big the
$\ln(a)$ got, we define the number $\mathcal{N}$ of e-foldings
after first horizon crossing until the end of inflation. Of course
$\mathcal{N}$ differs for each mode, but not much for the window
of observable modes. It's about $50$, and this is a reasonable
estimate for the value of $\ln(a)$.

At time $t_k$ the fluctuation with physical wave number $k_{\rm
phys}(t)\!=\!k/a(t)$ crosses the horizon for the first time, hence
$k_{\rm phys}(t_k)\!=\!H(t_k)$. Because the expansion rate $H$ is
constant during de Sitter inflation, this mode has comoving wave
number\beeq k\!=\!H a(t_k)\!=\!He^{Ht_k}\;
.\label{MODEcrossing}\eneq Using
Eqs.~(\ref{powerspectrum})-(\ref{MODEcrossing}) we find the
spectral index for the mode $k$ at the time of first horizon
crossing, up to $\mathcal{O}(\lambda^3)$, as\be &&\hspace{-0.4cm}
n_\varphi(t_k, k)\!=\!\frac{1}{\Delta^2_\varphi(t_k,
k)}\!\left[k\frac{\partial}{\partial
k}\!+\!a(t_k)\frac{\partial}{\partial
a(t_k)}\right]\!\Delta^2_\varphi(t_k, k)\\
&&\hspace{-0.3cm}=\!-\frac{\lambda}{12\pi^2}\Bigg[\!\ln(a)\!-\!\frac{1}{3}\!+\!\frac{a^{-3}}{3}
\!+\!\frac{k^2}{H^2}\Bigg(\!\frac{1}{5}\!-\!a^{-2}\!+\!a^{-3}\!+\!\mathcal{O}(a^{-5})\!\!\Bigg)
\!-\!\frac{k^4}{H^4}\Bigg(\!\frac{3}{70}\!+\!\frac{\ln(a)}{a^4}\!+\!\mathcal{O}(a^{-4})\!\!\Bigg)
\!+\!\mathcal{O}\Big(\!\frac{k^6}{H^6}\!\Big)\!\Bigg]\nonumber\\
&&\hspace{-0.3cm}+\frac{\lambda^2}{216\pi^4}
\Bigg[\!\ln^3(a)\!+\!\frac{3\ln^2(a)}{16}\!-\!\left(\!\!\frac{185}{32}\!-\!\frac{\pi^2}{2}
\!-\!\frac{27}{8}\ln\!\Big(\frac{2\mu}{H}\!\Big)\!\!\right)\!\ln(a)\!+\!\frac{817}{288}
\!+\!\frac{5\pi^2}{48}\!+\!\frac{9}{8}\ln^2\!\Big(\frac{2\mu}{H}\!\Big)\!-\!3\ln\!\Big(\frac{2\mu}{H}\!\Big)\nonumber\\
&&\hspace{-0.3cm}-3\zeta(3)
\!+\!\frac{9}{2}\sum_{p=2}^\infty\!\sum_{n=1}^{p-1}\frac{1}{n(p-n)p(p+3)}\!+\!\frac{81}{16a^2}
\!-\!\frac{\ln^2(a)}{4a^3}\!-\!\Big(\frac{19}{96}\!+\!\frac{\pi^2}{4}\!\Big)\frac{\ln(a)}{a^3}\!+\!\mathcal{O}(a^{-3})\nonumber\\
&&\hspace{-0.3cm}+\frac{k^2}{H^2}\!\Bigg(\!\!\Bigg(\!\frac{13283}{6000}\!-\!\frac{\pi^2}{10}
\!\Bigg)\!\ln(a)\!+\!\frac{13283}{12000}\!-\!\frac{\pi^2}{20}
\!+\!\frac{\mathcal{C}_2}{8}\!-\!\frac{3\ln^3(a)}{2a^2}\!-\!\frac{3\ln^2(a)}{2a^2}\!+\!\frac{15\ln(a)}{2a^2}
\!+\!\mathcal{O}(a^{-2})\!\Bigg)\nonumber\\
&&\hspace{-0.3cm}-\frac{k^4}{H^4}\!\left(\!\!\left(\!\frac{548167}{686000}
\!-\!\frac{3\pi^2}{70}
\!\right)\!\ln(a)\!+\!\frac{589327}{2744000}\!-\!\frac{3\pi^2}{280}
\!+\!\frac{\mathcal{C}_4}{4}\!+\!\frac{3\ln^2(a)}{10a^2}\!+\!\frac{\ln(a)}{10a^2}\!+\!\mathcal{O}(a^{-2})\!\!\right)
\!+\!\mathcal{O}\Big(\frac{k^6}{H^6}\!\Big)\!\Bigg]\; ,\nonumber\\
&&\hspace{10cm}\label{nphifinal} \ee where the terms that depend
on the scale factor $a$ in Eq.~(\ref{nphifinal}) must be evaluated
at $t\!=\!t_k$. To get a reasonable estimate of $n_\varphi(t, k)$
after the end of inflation, $\ln(a)$ may be taken as the number of
e-foldings $\mathcal{N}\sim 50$. We infer from
Eq.~(\ref{nphifinal}) that the dominant one-loop quantum effect
red tilts the spectrum
---although at two-loop order the effect contributes as a blue
tilt. Therefore, the amplitudes of fluctuations grow slightly
towards the larger scales. How much the spectral index $n_\varphi$
changes as the scale varies is measured by the running of the
index\beeq \alpha_\varphi(t, k)\!=\!\frac{d\,n_\varphi(t,
k)}{d\ln(k)}\; .\label{running}\eneq At the time of first horizon
crossing \be &&\hspace{-0.5cm}\alpha_\varphi(t_k,
k)\!=\!-\frac{\lambda}{12\pi^2}\!\left[1\!-\!a^{-3}
\!+\!\frac{k^2}{H^2}\Big(\frac{2}{5}\!-\!a^{-3}\!+\!\frac{3}{5a^{5}}\Big)\!-\!\frac{k^4}{H^4}\Big(\!\frac{6}{35}
\!+\!a^{-4}\!-\!\frac{3}{5a^{5}}\!-\!\frac{4}{7a^{7}}\!\Big)
\!+\!\mathcal{O}\Big(\frac{k^6}{H^6}\Big)\!\right]\nonumber\\
&&\hspace{-0.4cm}+\frac{\lambda^2}{72\pi^4}
\Bigg[\!\ln^2(a)\!+\!\frac{\ln(a)}{8}\!-\!\frac{185}{96}
\!+\!\frac{\pi^2}{6}\!+\!\frac{9}{8}\ln\!\Big(\frac{2\mu}{H}\!\Big)\!-\!\frac{27}{8a^{2}}\!+\!\frac{\ln^2(a)}{4a^3}
\!+\!\Big(\!\frac{1}{32}\!+\!\frac{\pi^2}{4}\!\Big)\frac{\ln(a)}{a^3}\!+\!\mathcal{O}(a^{-3})\nonumber\\
&&\hspace{-0.4cm}+\frac{k^2}{H^2}\!\left(\!\!\left(\!\frac{13283}{9000}
\!-\!\frac{\pi^2}{15} \!\right)\!\!\Big(\!\!\ln(a)\!+\!1\!\Big)
\!+\!\frac{\mathcal{C}_2}{12}\!-\!\frac{3\ln^2(a)}{2a^2}
\!-\!\frac{\ln(a)}{a^2}\!+\!\mathcal{O}(a^{-2})\!\!\right)\!-\!\frac{k^4}{H^4}\!\Bigg(\!\!\!\left(\frac{548167}{514500}
\!-\!\frac{2\pi^2}{35}
\!\right)\!\ln(a)\nonumber\\
&&\hspace{2.5cm}+\frac{568747}{1029000}\!-\!\frac{\pi^2}{35}
\!+\!\frac{\mathcal{C}_4}{3}\!+\!\frac{\ln^2(a)}{5a^2}
\!+\!\frac{4\!\ln(a)}{15a^2}\!+\!\mathcal{O}(a^{-2})\!\!\Bigg)
\!+\!\mathcal{O}\Big(\!\frac{k^6}{H^6}\!\Big)\!\Bigg]\!\!+\!\mathcal{O}(\lambda^3)\;
.\label{rng}\ee The terms that depend on $a$ in Eq.~(\ref{rng})
are evaluated at $t\!=\!t_k$. At one-loop order the leading terms
of $\alpha_\varphi$ are constants at zeroth, quadratic and quartic
order in $k/H$. After the first horizon crossing, inflationary
expansion enhances this result only at two-loop order. Thus, the
running of the spectral index towards larger scales is indeed
slight.

\section{Conclusions}
\label{sec:concl}

We have solved the one- and two-loop corrected, linearized
effective field equation for a massless, minimally coupled scalar
endowed with a quartic self-interaction in locally de Sitter
background for a state released in Bunch-Davies vacuum at time
$t=0$. The solution is the quantum corrected mode function of the
scalar at one- and two-loop orders. The quantum corrections can be
expanded in powers of $k/H$ and are proportional to $u(\infty,
k)$; the late-time limit of the tree-order mode function. See
Eqs.~(\ref{exactPhi1})-(\ref{power1}) and (\ref{Phi2}). Note that,
the shifts linear in $k/H$ are absent in both loop corrections. At
one-loop order, the quantum shift that is zero order in $k/H$
grows proportional to $(Ht)^2$ whereas the second and higher order
shifts in $k/H$ are constants, in leading order. At two-loop
order, on the other hand, the quantum shift that is zero order in
$k/H$ grows proportional to $(Ht)^4$, whereas the second and
higher order shifts in $k/H$ grow proportional to $(Ht)^2$, in
leading order.

We have also computed the scalar's power spectrum
$\Delta^2_\varphi(t, k)$. Quantum corrections induce scale
dependent shifts in even powers of $k/H$ to the scale-free
tree-order late-time spectrum; see Eq.~(\ref{powerspectrum}). At
one loop order, the quantum shift that is zero order in $k/H$
grows proportional to $(Ht)^2$ whereas the shifts quadratic and
quartic in $k/H$ are constants, in leading order. At two-loop
order, the quantum shift that is zero order in $k/H$ grows
proportional to $(Ht)^4$, whereas the shifts quadratic and quartic
in $k/H$ grows proportional to $(Ht)^2$, in leading order. The
spectral index $n_\varphi(t, k)$ vanishes at tree-order; see
Eq.~(\ref{nphifinal}). It is purely quantum mechanical in origin.
At one-loop order, the quantum shift that is zero order in $k/H$
grows proportional to $Ht$ whereas the shifts quadratic and
quartic in $k/H$ are constants, in leading order. At two-loop
order, on the other hand, the quantum shift that is zero order in
$k/H$ grows proportional to $(Ht)^3$, whereas the shifts quadratic
and quartic in $k/H$ grows proportional to $Ht$, in leading order.
The running of the spectral index $\alpha_\varphi(t, k)$ is given
in Eq.~(\ref{rng}). We conclude that the quantum effects red-tilts
the spectrum slightly. Hence, the amplitudes of fluctuations grow
towards the larger scales.

Because we cannot even measure the amplitude of our spectator
scalar, currently there isn't any data on either its spectral
index or its tilt. Can, {\it in principle}, the quantum shifts we
calculated in the spectrum be detected, {\it assuming} we can
measure the tree-order power spectrum for the scalar we consider?
The answer to the question depends upon how sensitively we can
measure the spectrum and what the value of the self-coupling
$\lambda$ is. One also needs to be able to distinguish between any
loop correction and a change in the inflaton~potential.

A phenomenologically more interesting application of the
computation presented in this paper would be tying it to Higgs
inflation \cite{BeSh}. For Higgs inflation one needs the scalar
background not to be $\varphi\!=\!0$. One also needs a strong
conformal coupling
--- so the scalar would not any longer be even approximately
minimally coupled.

\begin{appendix}
\section{Evaluating the functional $\xi(\alpha)$} \label{App:xi}
In Sec.~\ref{sec:Orderlamdasquarred}, while evaluating
$\mathcal{I}_1(\eta, k)$, we defined the functional $\xi(\alpha)$
in Eq.~(\ref{xialphafirst}) as\be \xi(\alpha) \!&\!\equiv
\!\!&\!\int_0^1\!\!\!dz\, z \sin(\alpha z) \ln(1\!-\!
z^2)\!=\!\!\int_0^1 \!\!dz\, z \sin(\alpha z) \left[\ln(1\!-\!
z)\!+\!\ln(1\!+\!z)\right]\; .\ee In this appendix we evaluate the
$\xi(\alpha)$. The first step is to make the change of variable
$y\!=\!1\!-\!z$ for the integral involving $\ln(1\!-\!z)$ and
$y\!=\!1\!+\!z$ for the other integral involving $\ln(1\!+\!z)$.
Then, we find \be \xi(\alpha)\!=\!\!\int_0^2\!\!dy (1\!-\!y)
\ln(y)\sin(\alpha(1\!-\!y))\!=\!-\frac{d}{d\alpha}\int_0^2\!\!dy
\ln(y) \cos(\alpha\!-\!\alpha y)\; .\ee Expanding out the cosine
term and changing the variable from $y$ to $x\!=\!\alpha y$, we
obtain \be\xi(\alpha)\!=\!-\frac{d}{d\alpha}
\left\{\frac{1}{\alpha}\left[\cos(\alpha)\!\!\int_0^{2\alpha}\!\!dx\ln(x)\cos(x)
\!+\!\sin(\alpha)\!\!\int_0^{2\alpha}\!\!dx\ln(x)\sin(x)\!-\!2\ln(\alpha)\sin(\alpha)\right]\right\}
\nonumber\ee Finally, recognizing the sine integral ${\rm
si}(\alpha)$ and cosine integral ${\rm ci}(\alpha)$ in the form
\be
   \int_0^{2\a} \!\!dx \ln(x) \cos(x)\!\!&=&\!\!\ln(2\a) \sin(2\a) \!-\! \frac{\pi}{2} \!-\! {\rm si}(2\a)\; ,\\
   \int_0^{2\a} \!\!dx \ln(x) \sin(x)\!\!&=&\!\!-\ln(2\a) \cos(2\a) \!-\! \gamma \!+\! {\rm ci}(2\a) \; ,\ee
where\be {\rm si}(\a) \!\!&\equiv&\!\! - \!\int_\a^{\infty} \!\!dt
{\sin(t) \over t} = -\frac{\pi}2
\!+\!\!\int_0^\a \!\!dt {\sin(t) \over t} \; , \label{siint}\\
{\rm ci}(\a) \!\!&\equiv&\!\! -\!\int_\a^{\infty} \!\!dt {\cos(t)
\over t} = \gamma \!+\! \ln(\a) \!+\!\!\int_0^\a \!\!dt {\cos(t)
\!-\!1 \over t} \label{ciint}\; , \ee with $\gamma \approx .577$
being the Euler's constant, and using the identities\beeq
\frac{d}{d \a} {\rm si(\a)}=\frac{\sin(\a)}{\a} \; , \hskip 0.5cm
\frac{d}{d \a} {\rm ci(\a)}=\frac{\cos(\a)}{\a}\; ,\eneq we
find\be \xi(\alpha)\!=\!\frac2{\alpha^2}
\sin(\alpha)\!\!&-&\!\!\frac1{\alpha^2} [\cos(\alpha)\!+\!\alpha
\sin(\alpha)]\!\left[{\rm si}(2 \alpha)\!+\!\frac{\pi}{2}
\right]\nonumber\\\!&+&\!\!\frac1{\alpha^2} [\sin(\alpha)\!-\!
\alpha \cos(\alpha)]\!\left[ {\rm ci}(2 \alpha)\!-\!
\ln\Bigl(\frac{\alpha}2\Bigr)\!-\!\gamma \right]\; . \ee The
functional $\xi(\alpha)$ also appears in the evaluations of the
integrals $I_2(\eta, k)$, $I_3(\eta, k)$, $I_4(\eta, k)$ and
$I_5(\eta, k)$; see Eqs.~(\ref{I2zetaalpha}), (\ref{st4}),
(\ref{WalphaI4}) and (\ref{WalphaI5}), respectively. Note that the
following expansions of the ${\rm si}(\a)$ and ${\rm ci}(\a)$ are
useful through out the paper \be
  {\rm si}(\a)\!+\!\frac{\pi}{2} \!\!&=&\!\! \sum_{n=1}^\infty
                 \frac{(-1)^{n+1}\a^{2n-1}}{(2n-1)\,(2n-1)!}\; ,\label{sipowerseri}\\
  {\rm ci}(\a)\!-\!\ln(\a)\!-\!\gamma \!\!&=&\!\! \sum_{n=1}^\infty
                 \frac{(-1)^{n}\a^{2n}}{2n\,(2n)!} \; .\label{cipowerseri}\ee

\section{Integrating $\mathcal{I}_n(\eta, k)$ for $2\leq n\leq 6$}
\label{App:In} In Sec.~\ref{sec:Orderlamdasquarred}, we obtained
the integro-differential equation~(\ref{IntegroPhi2}) for the
two-loop correction $\Phi_2(\eta, k)$ to the scalar mode function.
In the source term of the equation we defined integrals
$\mathcal{I}_n(\eta, k)$ (Eq.~(\ref{In})) and evaluated
$\mathcal{I}_1(\eta, k)$ through equations
(\ref{I1first})-(\ref{I1exp}). The evaluation of
$\mathcal{I}_2(\eta, k)$ is similar to the one for
$\mathcal{I}_1(\eta, k)$. Defining $\beta\equiv
\ln\left(\frac{He^{\frac34}}{2\mu}\right)$, we write \beeq
\mathcal{I}_2(\eta, k) = \frac{i H^2 \b}{2^9 \,\pi^6 \,a^2}\!\!
\int\!d^4x' a'^2\!\left\{\!\partial^2\!\left[\frac{\ln(\mu^2
\Delta x^2_{ \scriptscriptstyle ++})}{\Delta
x^2_{\scriptscriptstyle ++}} - \frac{ \ln(\mu^2 \Delta
x^2_{\scriptscriptstyle +-})}{\Delta x^2_{\scriptscriptstyle +-}}
\right]\!\right\} u(\eta',k) e^{-i \vec{k} \cdot (\vec{x}
-\vec{x}\,')}\nonumber\label{I2zetaalpha}\; . \eneq Following the
steps from Eq.~(\ref{logIdder}) to Eq.~(\ref{rad}) we obtain\be
\mathcal{I}_2(\eta, k)\!=\! \frac{-H^2\b}{2^7\,\pi^4}\,
\frac{\left(\partial_0^2 \!+\! k^2\right)^2}{a^2k^3} \!\!
\int_{\eta_i}^{ \eta} \!\! d\eta' a'^2 u(\eta',k)\Bigg\{\!\!\left[
\ln(\mu\D\eta)\!-\!\frac{1}{2}\right]\!\!\Big[\sin(\a)\!-\!\a\cos(\a)\Big]
\!+\!\frac{\a^2}{2}\,\xi(\a)\!\Bigg\}\nonumber \ee Applying the
derivatives using Eq.~(\ref{derivativeONintegral}) yields\be
&&\hspace{-1.2cm}\mathcal{I}_2(\eta, k)\!=\!
\frac{-H^2\b}{2^7\,\pi^4} \frac{\left(\partial_0^2 +
k^2\right)}{a^2k}\!\! \int_{\eta_i}^{ \eta}\!\!d\eta' a'^2
u(\eta',k)\Bigg\{\!\Big[{\rm ci}(2\a)
\!-\!\ln\left(2\a\right)\!-\!\g\!+\!
2\ln(2\mu\D\eta)\Big]\sin(\a)\nonumber\\&&\hspace{9.75cm}-\!\left[{\rm
si}(2\a)\!+\!\frac{\pi}{2}\right]\cos(\a) \Bigg\}\nonumber\\
&&\hspace{-1cm}=\!\frac{-H^2\b}{2^6\,\pi^4} \,
a^{-2}\Bigg\{\!k\!\!\int_{\eta_i}^{ \eta}\!\!\!\!d\eta' a'^2
u(\eta',k)\ln(2\mu\D\eta)\sin(\a)\!+\!\dd_0\!\!\int_{\eta_i}^{
\eta}\!\!\!\!d\eta' a'^2
u(\eta',k)\ln(2\mu\D\eta)\cos(\a)\!\Bigg\} \; .\ee We expand the
logarithms using Eq.~(\ref{logexpansion}) and evaluate the
integrals. Taking the necessary derivative and expressing the
final result in powers of $k/a$ we finally find\be
&&\hspace{-0.5cm}\mathcal{I}_2(\eta, k)\!=\!\frac{\b}{2^{6}
\,\pi^4}\frac{H}{\sqrt{2k^3}}
\Bigg\{\!H^2\!\!\left[\!\ln(a)\!-\!\ln\!\Big(\!\frac{2\,\mu}{He}\!\Big)
\!+\!\!\sum_{n=1}^\infty\frac{a^{-(n+1)}}{n\!+\!1}\right]\nonumber\\
&&\hspace{-0.15cm}-\frac{k^2}{a^2}\!\left[\frac{1}{2}\ln\!\Big(\!\frac{2\,\mu}{Ha}\!\Big)
\!+\!\frac{1}{2}\!-\!a^{-1}
\!-\!\frac{1}{2}\!\sum_{n=1}^\infty\frac{a^{-(n+1)}}{n\!+\!1}\right]\!-\!\frac{ik^3}{Ha^3}\!\left[\frac{a}{3}
\!+\!\frac{1}{3}\ln\!\Big(\!\frac{2\,\mu}{He}\!\Big)
\!-\!\frac{a^{-1}}{3}\!-\!\frac{1}{3}\!\sum_{n=1}^\infty\frac{a^{-(n+1)}}{n\!+\!1}\right]\nonumber\\
&&\hspace{2.5cm}+\frac{k^4}{H^2a^4}\!\left[\frac{a^2}{6}\!-\!\frac{1}{8}\ln(a)
\!+\!\frac{1}{8}\ln\!\Big(\!\frac{2\,\mu}{He}\!\Big)
\!-\!\frac{a^{-1}}{6}\!-\!\frac{1}{8}\!\sum_{n=1}^\infty\frac{a^{-(n+1)}}{n\!+\!1}\right]
\!+\mathcal{O}\Big(\frac{ik^5}{H^3}\Big)\!\Bigg\}\;
.\label{I2exp}\ee

To evaluate the third integral \be
\hspace{-1cm}&&\mathcal{I}_3(\eta, k)\!\equiv\! \frac{iH^2}{2^{11}
\,\pi^6\,a^2}\!\int\!\!d^4x' a'^2
\!\left\{\!\partial^2\!\!\left[\frac{\ln^2(\mu^2 \Delta x^2_{
\scriptscriptstyle ++})}{\Delta x^2_{\scriptscriptstyle ++}} \!-\!
\frac{ \ln^2(\mu^2 \Delta x^2_{\scriptscriptstyle +-})}{\Delta
x^2_{\scriptscriptstyle +-}} \right]\!\right\}\! u(\eta',k) e^{-i
\vec{k} \cdot (\vec{x} -\vec{x}\,')}\; ,\ee in the source
term~(\ref{STermIntegroDE}) of the integro-differential
equation~(\ref{diffeqmainPHI2}), we first use the identity\be
\frac{\ln^2(\mu^2 \Delta x^2)}{\Delta
x^2}=\frac{\dd^2}{12}\left[\ln^3(\mu^2\D x^2)\!-\!3\ln^2(\mu^2\D
x^2)\!+\!6\ln(\mu^2\D x^2)\right]\; .\label{logsqrid}\ee Then, the
usual cancelation between the $++$ and $+-$ terms outside
light-cone via Eq.~(\ref{id4}) converts the integral to the
form\be &&\hspace{-1cm}\mathcal{I}_3(\eta,
k)\!=\!\frac{-H^2}{2^{12} \,\pi^5}e^{-i \vec{k}
\cdot\vec{x}}\frac{\partial^4}{a^2}
\!\int_{\eta_i}^\eta\!\!\!d\eta' a'^2 u(\eta',k)\!\int\!\!d^3x'
e^{i \vec{k} \cdot \vec{x}\,'}
\Theta(\D\eta\!-\!r)\nonumber\\&&\hspace{3.5cm}\times\left\{\ln^2(\mu^2(\Delta\eta^2\!-\!r^2))\!-\!2\ln(\mu^2(\Delta\eta^2\!-\!r^2))\!-\!\frac{\pi^2}{3}\!+\!2
\!\right\}\; .\ee Evaluating the angular integrations using
Eq.~(\ref{angular}), passing the spatial phase factor through the
derivative using $\partial^4 e^{i \vec{k} \cdot \vec{x}} = e^{i
\vec{k} \cdot \vec{x}} \left(
\partial_0^2\!+\!k^2 \right)^2$ and making the usual change of
variable $r\equiv\Delta\eta z\equiv\frac{\alpha}{k}z$ yields \be
&&\hspace{-1.2cm}\mathcal{I}_3(\eta, k)\!=\!\frac{-H^2}{2^8
\,\pi^4}
\frac{(\partial^2_0\!+\!k^2)^2}{a^2k^3}\!\!\int_{\eta_i}^\eta
\!\!\!d\eta' a'^2
u(\eta',k)\Bigg\{\!\Bigg[\!\ln^2(\mu\Delta\eta)\!-\!\ln(\mu\Delta\eta)\!-\ln^2(2)
\!-\!\frac{\pi^2}{12}\!+\!\frac{1}{2}\Bigg]\nonumber\\&&\hspace{3.5cm}
\times\Big[\sin(\alpha)\!-\!\alpha\cos(\alpha)\Big]
\!+\!\!\left[\ln(2\mu\Delta\eta)\!-\!\frac{1}{2}\right]\!\alpha^2\xi(\alpha)
\!+\!\frac{\a^2}{4} \mathcal{W}(\alpha) \!\Bigg\}\; ,
\label{st4}\ee where we define a new functional\be
\mathcal{W}(\alpha)\!&=&\!\!\!\int_0^1 \!\!\!dz z \sin(\alpha
z)\ln^2\!\Big(\frac{1\!-\!z^2}{4}\Big)\label{Walpha}\\\!&=&\!\!\sum_{n=0}^\infty
\frac{(-1)^n\alpha^{2n+1}}{(2n+1)!}\!\int_0^1\!\!dz
z^{2n+2}\ln^2\!\Big(\frac{1\!-\!z^2}{4}\Big)\label{2n2}\\
\!&=&\!\!\left(\frac{104}{27}\!-\!\frac{\pi^2}{9}\right)\alpha
\!-\!\left(\frac{1576}{3375}\!-\!\frac{\pi^2}{90}\right)\alpha^3\!+\!\left(\frac{21946}{1157625}
\!-\!\frac{\pi^2}{2520}\right)\a^5\!+\!O(\alpha^7)\; .\ee See
Appendix~\ref{App:w} for the evaluation of the
$\mathcal{W}({\alpha})$. Applying $\frac{\dd_0^2+k^2}{k^2}$ in
Eq.~(\ref{st4}) yields \be &&\hspace{-0.7cm}\mathcal{I}_3(\eta,
k)\!=\!\frac{- H^2}{2^8 \,\pi^4}
\frac{(\partial^2_0\!+\!k^2)}{a^2k} \!\!\int_{\eta_i}^\eta\!\!\!
d\eta' a'^2 u(\eta',k)\Bigg\{\!\sin(\a)\Big\{
\frac{\rm{si}(2\a)\!+\!\frac{\pi}{2}}{\a}\!+\!2\ln^2(2\mu\D\eta)
\!-\!\frac{\pi^2}{6}\nonumber\\
&&\hspace{0.7cm}+\Big[\rm{ci}(2\a)\!-\!\ln(2\a)\!-\!\gamma\Big]\!
\Big[2\ln(2\mu\D\eta)\!+\!1\!-\!\frac{1}{\a^2}\Big]\!-\!\frac{4}{\a^2}\!\Big\}
\!+\!\cos(\a)\Big\{\frac{\rm{ci}(2\a)\!-\!\ln(2\a)\!-\!\g\!+\!2}{\a}\nonumber\\
&&\hspace{0.7cm}-\!\left[\rm{si}(2\a)
\!+\!\frac{\pi}{2}\right]\!\!\Big[2\ln(2\mu\D\eta)\!+\!1\!-\!\frac{1}{\a^2}\Big]\!\Big\}
\!+\!\frac{\a^2}{4}\mathcal{W}(\a)\!+\!\frac{\partial^2_0}{k^2}\Big[\frac{\a^2}{4}\mathcal{W}(\a)\Big]\!\Bigg\}\;
. \label{I3middle}\ee Then, we use Eqs.~(\ref{logexpansion}),
(\ref{sipowerseri})-(\ref{cipowerseri}) in Eq.~(\ref{I3middle}),
evaluate the integrals over the past light-cone and take the
remaining derivatives. Expressing the result in powers of $k/a$, we
find \be &&\hspace{-0.5cm}\mathcal{I}_3(\eta, k)\!=\!\frac{-
1}{2^8
\,\pi^4}\frac{H}{\sqrt{2k^3}}\Bigg\{\!H^2\!\Bigg[2\!\ln^2\!\!\Big(\!\frac{2\mu}{Ha}\!\Big)
\!-\!4\!\ln\!\Big(\!\frac{2\mu}{Ha}\!\Big)\!+\!\frac{\pi^2}{3}
\!+\!4\!\sum_{n=1}^\infty\!\frac{na^{-(n+1)}}{(n\!+\!1)^2}\!-\!
4\!\ln\!\Big(\frac{2\mu}{H}\!\Big)\!\!\sum_{n=1}^\infty\!\frac{a^{-(n+1)}}{n\!+\!1}\nonumber\\
&&\hspace{-0.5cm}+2\!\sum_{p=2}^{\infty}\sum_{n=1}^{p-1}\!\frac{a^{-(p+1)}}{n(p-n)}[p\!-\!(p\!-\!1)a]\Bigg]
\!\!+\!\frac{k^2}{a^2}\!\Bigg[\!\ln^2\!\Big(\frac{2\mu}{Ha}\!\Big)\!+\!2\!\ln\!\Big(\frac{2\mu}{H}\Big)\!
\!-\!\frac{\pi^2}{18}\!-\!\frac{16}{9}\!+\!a^{-1}\!\Big[2\!-\!4\!\ln\!\Big(\frac{2\mu}{H}\Big)\Big]\nonumber\\
&&\hspace{-0.5cm}+a^{-2}\Big[\frac{31}{6}\!-\!\ln\!\Big(\frac{2\mu}{H}\!\Big)\Big]\!
\!-\!a^{-3}\Big[\frac{32}{27}\!+\!\frac{2}{3}\ln\!\Big(\!\frac{2\mu}{H}\!\Big)\Big]
\!+\!\frac{2}{3}\!\sum_{n=3}^\infty\!\frac{(3n^2\!-\!2)a^{-(n+1)}}{n(n\!+\!1)^2}
-2\ln\!\Big(\frac{2\mu}{H}\Big)\!\!\sum_{n=3}^\infty\!\frac{a^{-(n+1)}}{n\!+\!1}\nonumber\\
&&\hspace{-0.5cm}+\!\sum_{p=3}^{\infty}\sum_{n=1}^{p-1}\frac{1}{n(p\!-\!n)}\Bigg\{\!
\frac{4}{(p\!-\!2)(p+\!1)}-\!a^{-(p+1)}\frac{p}{3}\!\left(\!\frac{p+\!5}{p+\!1}
\!-\!a\,\frac{3\,(p\!+\!3)}{p}\!+\!a^3\frac{2\,(p\!-\!1)}{p\!-\!2}\!\right)\!\!\Bigg\}\Bigg]\nonumber\\
&&\hspace{-0.5cm}+\frac{ik^3}{Ha^3}\!\Bigg[\frac{4}{3}\ln\!\Big(\frac{2\mu}{He}\!\Big)a\!-\!\frac{440\ln(a)}{63}
\!+\!\frac{\pi^2}{9}\!+\!\frac{92}{189}\!+\!\frac{2}{3}\ln^2\!\Big(\frac{2\mu}{H}\!\Big)
\!-\!\frac{4}{3}\ln\!\Big(\frac{2\mu}{H}\!\Big)
\!+\!8\,{a^{-1}}\!+\!\frac{4}{3}\!\sum_{n=0}^\infty\!\frac{na^{-(n+1)}}{(n\!+\!1)^2}\nonumber\\
&&\hspace{-0.5cm}-\frac{4}{3}\!\ln\!\Big(\!\frac{2\mu}{H}\!\Big)\!\!\sum_{n=0}^\infty\!\frac{a^{-(n+1)}}{(n\!+\!1)}
\!+\!\sum_{p=4}^{\infty}\!\sum_{n=1}^{p-1}\!\frac{1}{n(p\!-\!n)}\Bigg\{\!\frac{4}{(p\!-\!3)(p\!-\!2)}
\!+\!\frac{2}{3}a^{-(p-2)}p\left(\!\frac{p\!+\!1}{p-2}
\!-\!a\frac{(p\!-\!1)}{p-3}\!\right)\!\!\Bigg\}\!\Bigg]\nonumber\\
&&\hspace{-0.5cm}+\frac{k^4}{H^2a^4}\!\Bigg[a^2\Big[\frac{13}{18}
\!-\!\frac{2}{3}\ln\!\Big(\frac{2\mu\!}{H}\Big)\Big]\!+\!a\!+\!\frac{1177}{900}\!+\!\frac{\pi^2}{40}
\!+\!\frac{1}{2}\!\ln\!\Big(\!\frac{\!2\mu}{Ha}\!\Big)\!-\!\frac{1}{4}\!\ln^2\!\Big(\!\frac{2\mu}{Ha}\!\Big)
\!+\!a^{-1}\!\Big[\frac{2}{3}\!\ln\!\Big(\frac{2\mu}{H}\!\Big)\!-\!\frac{20}{3}\Big]\nonumber\\
&&\hspace{-0.5cm}+a^{-2}\Big[\frac{407}{120}
\!+\!\frac{1}{4}\!\ln\!\Big(\frac{2\mu}{H}\!\Big)\Big]
\!+\!a^{-3}\Big[\frac{1}{6}\!\ln\!\Big(\frac{2\mu}{H}\!\Big)\!-\!\frac{187}{225}\Big]
\!\!-\!\frac{497}{720}a^{-4}\!+\!\frac{143}{900}a^{-5}\nonumber\\
&&\hspace{-0.5cm}-\frac{1}{30}\!\sum_{n=3}^\infty\!\frac{(15n^4\!
+\!75n^3\!+\!110n^2\!+\!20n\!-\!24)a^{-(n+1)}}{n(n\!+\!1)^2(n\!+\!2)(n\!+\!3)}
\!+\!\frac{1}{2}\!\ln\!\Big(\frac{2\mu}{H}\!\Big)\!\!\sum_{n=3}^\infty\!\frac{a^{-(n+1)}}{n\!+\!1}\nonumber\\
&&\hspace{-0.5cm}-\sum_{p=5}^{\infty}\sum_{n=1}^{p-1}\frac{1}{n(p\!-\!n)}\Bigg\{\!
\frac{2(p^2\!+\!p\!+\!4)}{(p\!-\!4)(p\!-\!3)(p\!-\!2)(p+\!1)}\nonumber\\
&&\hspace{-0.5cm}-a^{-(p+1)}p\!\left(\!\frac{p+\!9}{60(p+\!1)}
\!-\!a\,\frac{(p\!+\!7)}{12\,p}\!+\!a^3\frac{(p\!+\!3)}{3(p\!-\!2)}\!-\!a^4\frac{2(p\!+\!1)}{3(p\!-\!3)}
\!+\!a^5\frac{2(p\!-\!1)}{5(p\!-\!4)}\!\right)\!\!\Bigg\}\Bigg]
\!+\!\mathcal{O}(\frac{ik^5}{H^3})\!\Bigg\}\; .\label{I3exp}\ee

The fourth integral in  source term~(\ref{STermIntegroDE}) \beeq
\mathcal{I}_4(\eta, k)\!=\!\frac{i H^4}{2^9
\,\pi^6}\frac{1}{a}\!\int\!\! d^4x' a'^3 \!\Bigg\{
\frac{\ln^2(\frac{\sqrt{e}H^2}{4} \Delta x^2_{ \scriptscriptstyle
++})}{\Delta x^2_{\scriptscriptstyle ++}} \!-\! \frac{
\ln^2(\frac{\sqrt{e}H^2}{4} \Delta x^2_{\scriptscriptstyle
+-})}{\Delta x^2_{\scriptscriptstyle +-}}\!\Bigg\} u(\eta',k)
e^{-i \vec{k}\cdot (\vec{x} -\vec{x}\,')}\; ,\label{I4middle}\eneq
is evaluated similarly to the third one. We use
Eq.~(\ref{logsqrid}) in Eq.~(\ref{I4middle}) and obtain\be
&&\hspace{-1.5cm}\mathcal{I}_4(\eta,
k)\!=\!\frac{-H^4}{2^{10}\pi^5}e^{-i
\vec{k}\cdot\vec{x}}\frac{\partial^2}{a}\!\!\int_{\eta_i}^\eta
\!\!\!d\eta' a'^3 u(\eta',k)\!\!\int\!\!d^3x\,'e^{i
\vec{k}\cdot\vec{x}\,'}
\Theta(\Delta\eta-r)\nonumber\\
&&\hspace{3.3cm}\times\Bigg[\!\ln^2\!\Big(\!\frac{H^2(\!\Delta\eta^2\!-\!r^2)}{4}\Big)
\!-\!\ln\!\Big(\!\frac{H^2(\Delta\eta^2\!-\!r^2)}{4}\Big)\!+\!\frac{5}{4}\!-\!\frac{\pi^2}{3}\Bigg]\;
.\label{intfournewform}\ee Evaluating the angular integrations
using Eq.~(\ref{angular}), passing the spatial phase factor
through the derivative using $\partial^2 e^{i \vec{k} \cdot
\vec{x}}\!=\!-e^{i \vec{k} \cdot \vec{x}} \left(
\partial_0^2\!+\!k^2\right)$ and then making the change of
variable $r\!\equiv\!\frac{\alpha}{k}z$ reduces
Eq.~(\ref{intfournewform}) to the form \be
&&\hspace{-1.3cm}\mathcal{I}_4(\eta, k)\!=\!\frac{H^4}{2^6
\,\pi^4}
\frac{(\partial^2_0+k^2)}{ak^3}\!\!\!\int_{\eta_i}^\eta\!\!\!\!d\eta'
a'^3u(\eta',k)\Bigg\{\!\Bigg[\!\ln^2\!\Big(\!\frac{H\Delta\eta}{2}\!\Big)
\!-\!\frac{1}{2}\!\ln\!\Big(\!\frac{H\Delta\eta}{2}\!\Big)
\!-\!\ln^2(2)\!-\!\frac{\pi^2}{12}\!+\!\frac{5}{16}\Bigg]\nonumber\\
&&\hspace{3cm}\times\Big[\!\sin(\alpha)\!-\!\alpha\cos(\alpha)\Big]
\!+\!\left[\ln\!\Big(\!H\Delta\eta\!\Big)\!-\!\frac{1}{4}\right]\alpha^2\xi(\alpha)
\!+\!\frac{\a^2}{4} \mathcal{W}(\alpha) \!\Bigg\}\;
.\label{WalphaI4} \ee When the derivatives are applied to the
integral using Eq.~(\ref{derivativeONintegral}),
$\mathcal{I}_4(\eta, k)$ becomes\be
&&\hspace{-0.5cm}\frac{H^4}{2^6
\,\pi^4}\frac{1}{ak}\!\!\int_{\eta_i}^\eta\!\!\!\!d\eta' a'^3
u(\eta',k)\Bigg\{\!\!\cos(\a)\!\Bigg\{\!\!\Big[\frac{1}{\a^2}\!-\!\frac{3}{2}\!-\!2\ln(\!H\!\D\eta)\!\Big]\!\!\left[{\rm
si}(2\a)\!+\!\frac{\pi}{2}\right]\!\!+\!\frac{{\rm
ci}(2\a)\!-\!\ln(2\a)\!-\!\g\!+\!2}{\a}\!\Bigg\}\nonumber\\
&&\hspace{1.5cm}+\sin(\a)\!\Bigg\{\frac{\rm{si}(2\a)\!+\!\frac{\pi}{2}}{\a}
\!-\!\Big[\frac{1}{\a^2}\!-\!\frac{3}{2}\!-\!2\ln(\!H\!\D\eta)\Big]\!\Big[\rm{ci}(2\a)\!-\!\ln(2\a)\!-\!\g\Big]
\!-\!\frac{4}{\a^2}\!-\!\frac{\pi^2}{6}\!+\!\frac{1}{8}\nonumber\\
&&\hspace{4cm}+\ln(\!H\!\D\eta)\!+\!2\ln^2(\!H\!\D\eta)\!\Bigg\}\!+\!\frac{\a^2}{4}\mathcal{W}(\a)
\!+\!\frac{\partial^2_0}{k^2}\Big[\frac{\a^2}{4}\mathcal{W}(\a)\Big]\!\Bigg\}
\; .\label{int4ara}\ee Then, we use Eqs.~(\ref{logexpansion}),
(\ref{sipowerseri})-(\ref{cipowerseri}) in Eq.~(\ref{int4ara}) and
integrate over the past light-cone. In powers of $k/a$ we have \be
&&\hspace{-0.6cm}\mathcal{I}_4(\eta, k)\!=\!\frac{1}{2^{6}
\,\pi^4}\frac{H}{\sqrt{2k^3}}
\Bigg\{\!H^2\!\Bigg[\!\ln^2(a)\!-\!\frac{\ln(a)}{2}
\!+\!\frac{\pi^2}{6}\!-\!\frac{7}{16}\!+\!\Big(\frac{\pi^2}{3}\!-\!\frac{33}{8}\Big)a^{-1}
\!+\!\Big(\frac{13}{16}\!-\!\frac{\pi^2}{6}\Big)a^{-2}\nonumber\\
&&\hspace{0cm}+\!\sum_{n=2}^\infty\frac{(3n\!-\!1)a^{-(n+1)}}{(n\!-\!1)n(n\!+\!1)^2}
\!+\!2\!\sum_{p=2}^{\infty}\!\sum_{n=1}^{p-1}\!\frac{1}{n(p\!-\!n)}\Bigg\{\!\frac{1}{(p\!+\!1)(p\!+\!2)}
\!-\!a^{-(p+1)}\left(\!\frac{1}{p\!+\!1}
\!-\!\frac{a^{-1}}{p\!+\!2}\!\right)\!\!\Bigg\}\!\Bigg]\nonumber\\
&&\hspace{0cm}-\frac{k^2}{a^2}\!\Bigg[\frac{\ln^2(a)}{2}\!+\!\frac{3\ln(a)}{4}
\!-\!\Big(\frac{629}{216}\!-\!\frac{\pi^2}{9}\Big)a\!+\!\frac{89}{96}
\!+\!\frac{5\pi^2}{36}\!-\!\Big(\frac{83}{48}
\!-\!\frac{\pi^2}{6}\Big)a^{-1}\!-\!\Big(\frac{11}{96}\!+\!\frac{\pi^2}{36}\Big)a^{-2}\nonumber\\
&&\hspace{0cm}-\frac{1}{6}\!\sum_{n=2}^\infty
\frac{(9n^2\!-\!21n\!-\!2)a^{-(n+1)}}{(n\!-\!1)n(n\!+\!1)^2(n\!+\!2)}
\!-\!\sum_{p=2}^{\infty}\!\sum_{n=1}^{p-1}\!\frac{1}{n(p\!-\!n)}
\Bigg\{\!\frac{p\!+\!3}{(p\!-\!1)(p\!+\!1)(p\!+\!2)}\nonumber\\
&&\hspace{0cm}-a^{-(p-1)}\!\left(\!\frac{2}{3(p\!-\!1)}
\!-\!\frac{a^{-2}}{p\!+\!1}\!+\!\frac{a^{-3}}{3(p\!+\!2)}\!\right)\!\!\Bigg\}\!\Bigg]
\!+\!\frac{ik^3}{Ha^3}\!\Bigg[\Big(\frac{15}{16}
\!-\!\frac{\pi^2}{18}\Big)a^2\!+\!\Big(\frac{\pi^2}{9}
\!-\!\frac{25}{24}\Big)a\!+\!\Big(\frac{\pi^2}{18}\!-\!\frac{79}{48}\Big)\nonumber\\
&&\hspace{0cm}+\frac{11a^{-1}}{18}\!+\!\frac{1}{3}\!\sum_{n=1}^\infty
\frac{(3n\!-\!1)a^{-(n+1)}}{(n\!+\!1)^2(n\!+\!2)(n\!+\!3)}
\!+\!\frac{2}{3}\!\sum_{p=3}^{\infty}\!\sum_{n=1}^{p-1}
\!\frac{1}{n(p\!-\!n)}\Bigg\{\!\frac{1}{(p\!-\!1)(p\!-\!2)}\nonumber\\
&&\hspace{0cm}-a^{-(p-2)}\!\left(\!\frac{1}{p\!-\!2}
\!-\!\frac{a^{-1}}{p\!-\!1}\!\right)\!\!\Bigg\}\Bigg]\!+\!\frac{k^4}{H^2a^4}
\!\Bigg[\Big(\frac{\pi^2}{45}\!-\!\frac{9403}{27000}\Big)a^3
\!+\!\Big(\frac{251}{432}\!-\!\frac{\pi^2}{18}\Big)a^2\!+\!\Big(\frac{\pi^2}{18}
\!-\!\frac{245}{432}\Big)a\nonumber\\
&&\hspace{0cm}+\frac{\ln^2(a)}{24}\!+\!\frac{13\ln(a)}{144}
\!+\!\frac{17\pi^2}{720}\!+\!\frac{10157}{86400}
\!-\!\frac{1}{4}\!\sum_{p=4}^{\infty}\!\sum_{n=1}^{p-1}\!\frac{p^3\!+\!2p^2\!
+\!3p\!+\!10}{n(p\!-\!n)(p\!-\!3)(p\!-\!2)(p\!-\!1)(p\!+\!1)(p\!+\!2)}\nonumber\\
&&\hspace{0cm}
+\Big(\frac{\pi^2}{72}\!-\!\frac{437}{960}\Big)a^{-1}\!+\!\Big(
\frac{839}{5760}\!-\!\frac{\pi^2}{720}\Big)a^{-2}\!-\!\frac{107\,a^{-3}}{3600}
\!-\!\frac{2789\,a^{-4}}{151200}\!+\!\frac{1651\,a^{-5}}{756000}\nonumber\\
&&\hspace{0cm}-\frac{1}{120}\!\sum_{n=5}^{\infty}\frac{(45n^4\!+\!30n^3\!+\!155n^2\!-\!470n\!-\!136)a^{-(n+1)}}
{(n\!-\!1)n(n\!+\!1)^2(n\!+\!2)(n\!+\!3)(n\!+\!4)}\Bigg]\!+\!\sum_{p=4}^{\infty}\!\sum_{n=1}^{p-1}
\!\frac{a^{-(p-3)}}{n(p\!-\!n)}\Bigg\{\!\frac{2}{5(p\!-\!3)}\nonumber\\
&&\hspace{4cm}-\frac{2a^{-1}}{3(p\!-\!2)}+\frac{a^{-2}}{3(p\!-\!1)}\!-\!\frac{a^{-4}}{12(p\!+\!1)}
\!+\!\frac{a^{-5}}{60(p\!+\!2)}\Bigg\}\!+\!\mathcal{O}\Big(\frac{ik^5}{H^3}\Big)\!\Bigg\}\;
.\label{I4exp}\ee

The fifth integral in the source term involves difference of two
cubic logarithms in the integrand, \beeq\mathcal{I}_5(\eta,
k)\!=\! \frac{-i H^6}{2^{10}3\,\pi^6}\!\!\int\!\! d^4x' a'^4
\!\Bigg\{ \!\!\ln^3\!\Big(\!\frac{\sqrt{e}}{4}H^2 \Delta x^2_{
\scriptscriptstyle ++}\!\Big) \!-\!
\ln^3\!\Big(\!\frac{\sqrt{e}}{4}H^2 \Delta x^2_{\scriptscriptstyle
+-}\!\Big)\!\Bigg\} u(\eta',k) e^{-i \vec{k}\cdot (\vec{x}
-\vec{x}\,')}\; .\label{int5}\eneq  Expanding the difference of
two cubes and using Eq.~(\ref{id4}) we find \be\mathcal{I}_5(\eta,
k)\!=\!\frac{ H^6}{2^9\pi^5}e^{-i\vec{k}\cdot
\vec{x}}\!\!\int_{\eta_i}^\eta\!\!\!d\eta' a'^4
u(\eta',k)\!\!\int\!\!d^3x'e^{i \vec{k} \cdot
\vec{x}\,'}\Theta(\D\eta\!-\!\D x)
\Bigg\{\!\!\ln^2\!\left(\!\!\frac{\sqrt{e}}{4}H^2(\D\eta^2\!-\!\Delta
x^2)\!\!\right)\!-\!\frac{\pi^2}{3}\!\Bigg\}\; .\nonumber\ee
Evaluating the angular integrations using Eq.~(\ref{angular}) and
making the change of variable $r\equiv\Delta\eta
z\equiv\frac{\alpha}{k}z$ yields\be
&&\hspace{-1.2cm}\mathcal{I}_5(\eta,
k)\!=\!\frac{H^6}{2^5\pi^4}\frac{1}{k^3}\!\!\int_{\eta_i}^\eta\!\!
d\eta' a'^4 u(\eta',k)\Bigg\{\!\!\left[
\ln^2\!\Big(\!\frac{H\D\eta}{2}\Big)
\!+\!\frac{1}{2}\ln\!\Big(\!\frac{H\D\eta}{2}\Big)\!-\!\ln^2(2)
\!+\!\frac{1}{16}\!-\!\frac{\pi^2}{12}\right]\nonumber\\
&&\hspace{2.7cm}\times\Big[\sin(\a)\!-\!\a\cos(\a)\Big]
\!+\!\left[\ln(H\D\eta)\!+\!\frac{1}{4}\right]\!\a^2\xi(\a)\!+\!\frac{\a^2}{4}\mathcal{W}(\a)\Bigg\}\;
.\label{WalphaI5} \ee We insert $\xi(\alpha)$ given in
Eq.~(\ref{xialphasecond}) into Eq.~(\ref{WalphaI5}) and regroup
the terms in the integrand, \be
&&\hspace{-1.4cm}\mathcal{I}_5(\eta,
k)\!=\!\frac{H^6}{2^5\pi^4}\frac{1}{k^3}\!\!\int_{\eta_i}^\eta\!\!
d\eta' a'^4
u(\eta',k)\Bigg\{\!\frac{\sin(\a)}{2}\!-\!\frac{1}{4}\Big[\rm{si}(2\a)\!+\!\frac{\pi}{2}\Big]
\!\Big[\cos(\a)\!+\!\a\sin(\a)\Big]\nonumber\\
&&\hspace{0.8cm}+\frac{1}{4}\Big[{\rm{ci}}(2\a)\!-\!\ln(2\a)\!-\!\g\!
+\!\frac{1}{4}\!-\!\frac{\pi^2}{3}\!+\!4\ln^2(H\Delta\eta)\Big]\Big[\sin(\a)\!-\!\a\cos(\a)\Big]\nonumber\\
&&\hspace{0.8cm}+\ln(H\Delta\eta)
\Big\{\!\sin(\a)\Big(\rm{ci}(2\a)\!-\!\ln(2\a)\!-\!\g\!-\a\Big[\rm{si}(2\a)\!+\!\frac{\pi}{2}\Big]
+\frac{5}{2}\Big)\nonumber\\
&&\hspace{1.8cm}-\cos(\a)\Big(\rm{si}(2\a)\!+\!\frac{\pi}{2}
\!+\!\a\Big[\rm{ci}(2\a)\!-\!\ln(2\a)\!-\!\g\!+\!\frac{1}{2}\Big]\!\Big)\!\Big\}\!+\!\frac{\a^2}{4}\mathcal{W}(\a)\!\Bigg\}\;
.\label{int5ara}\ee Using Eqs.~(\ref{logexpansion}),
(\ref{sipowerseri})-(\ref{cipowerseri}) in Eq.~(\ref{int5ara}) and
integrating over the past light-cone, we obtain\be
&&\hspace{-0.4cm}\mathcal{I}_5(\eta, k)=\!\frac{1}{2^{5}
\,\pi^4}\frac{H}{\sqrt{2k^3}}
\Bigg\{\!H^2\!\Bigg[\frac{\ln^3(a)}{9}\!-\!\frac{\ln^2(a)}{4}\!+\!\Big(\frac{\pi^2}{18}\!+\!\frac{5}{48}\Big)\ln(a)
\!+\!\frac{\pi^2}{54}\!-\!\frac{1}{96}\!-\!\frac{2\zeta(3)}{3}\nonumber\\
&&\hspace{-0.3cm}+\Big(\frac{33}{16}\!-\!\frac{\pi^2}{6}\Big)a^{-1}
\!+\!\Big(\frac{\pi^2}{12}\!-\!\frac{35}{96}\Big)a^{-2}\!-\!\Big(\frac{\pi^2}{54}\!+\!\frac{139}{1296}\Big)a^{-3}
\!+\!\frac{1}{3}\!\sum_{n=3}^\infty\frac{(13n\!+\!1)a^{-(n+1)}}{(n\!-\!2)(n\!-\!1)n(n\!+\!1)^3}\nonumber\\
&&\hspace{-0.3cm}+\!\sum_{p=2}^{\infty}\!\sum_{n=1}^{p-1}\!\frac{1}{n(p\!-\!n)}\Bigg\{\!\frac{2}{p(p\!+\!1)(p\!+\!2)(p\!+\!3)}
\!-\!a^{-p}\left(\!\frac{1}{3p}\!-\!\frac{a^{-1}}{p\!+\!1}
\!+\!\frac{a^{-2}}{p\!+\!2}\!-\!\frac{a^{-3}}{3(p\!+\!3)}\!\right)\!\!\Bigg\}\!\Bigg]\nonumber\\
&&\hspace{-0.3cm}+\frac{k^2}{a^2}\!\Bigg[\!\Big(\frac{13283}{54000}\!-\!\frac{\pi^2}{90}\Big)a^2
\!+\!\Big(\frac{\pi^2}{18}\!-\!\frac{629}{432}\Big)a\!
+\!\frac{\ln^3(a)}{18}\!+\!\frac{5\ln^2(a)}{24}\!+\!\Big(\frac{\pi^2}{36}\!+\!\frac{37}{96}\Big)\ln(a)\nonumber\\
&&\hspace{-0.3cm}+\frac{53\pi^2}{540}\!+\!\frac{2623}{43200}-\frac{\zeta(3)}{3}\!+\!\Big(
\frac{\pi^2}{36}+\frac{19}{480}\Big)a^{-1}\!-\!\Big(\frac{\pi^2}{72}\!+\!\frac{169}{2880}\Big)a^{-2}
\!+\!\Big(\frac{\pi^2}{540}\!-\!\frac{353}{12960}\Big)a^{-3}\nonumber\\
&&\hspace{-0.3cm}+\frac{863a^{-4}}{21600}\!-\!\frac{2491a^{-5}}{378000}\!+\!\frac{1}{30}\!\sum_{n=5}^\infty
\frac{(65n^3\!-\!190n^2\!-\!193n\!-\!58)a^{-(n+1)}}{(n\!-\!2)(n\!-\!1)n(n\!+\!1)^3(n\!+\!2)(n\!+\!3)}\nonumber\\
&&\hspace{-0.3cm}+\!\sum_{p=3}^{\infty}\!\sum_{n=1}^{p-1}\!\frac{1}{n(p\!-\!n)}
\Bigg\{\!\frac{p^2\!+\!5p\!+\!2}{(p\!-\!2)(p\!-\!1)p(p\!+\!1)(p\!+\!2)(p\!+\!3)}\!-\!a^{-(p-2)}\Bigg(\!\!\frac{2}{15(p\!-\!2)}
\!-\!\frac{a^{-1}}{3(p\!-\!1)}\nonumber\\
&&\hspace{-0.3cm}+\frac{a^{-2}}{6\,p}\!+\!\frac{a^{-3}}{6(p\!+\!1)}
\!-\!\frac{a^{-4}}{6(p\!+\!2)}\!+\!\frac{a^{-5}}{30(p\!+\!3)}\!\Bigg)\!\!\Bigg\}\!\Bigg]
\!+\!\frac{ik^3}{Ha^3}\!\Bigg[\!\Big(\!\frac{155}{1296}\!-\!\frac{\pi^2}{162}\Big)a^3\!+\!\Big(\frac{\pi^2}{36}
\!-\!\frac{15}{32}\Big)a^2\nonumber\\
&&\hspace{-0.3cm}+\Big(\frac{53}{48}\!-\!\frac{\pi^2}{18}\Big)a\!-\!\frac{\ln^3(a)}{27}\!-\!\frac{13\ln^2(a)}{108}
\!-\!\Big(\frac{\pi^2}{54}\!+\!\frac{329}{1296}\Big)\!\ln(a)\!+\!\frac{845}{7776}
\!-\!\frac{2\pi^2}{27}\!+\!\frac{2\zeta(3)}{9}\nonumber\\
&&\hspace{-0.3cm}+\frac{a^{-1}}{216}\!-\!\frac{233\,
a^{-2}}{2160}\!+\!\frac{41a^{-3}}{1080}\!+\!\frac{1}{9}\!\sum_{n=3}^\infty
\frac{(13\,n\!+\!1)a^{-(n+1)}}{(n\!+\!1)^3(n\!+\!2)(n\!+\!3)(n\!+\!4)}\nonumber\\
&&\hspace{-0.3cm}+\!\!\sum_{p=4}^{\infty}\!\sum_{n=1}^{p-1}
\!\frac{1}{n(p\!-\!n)}\Bigg\{\!\frac{2}{3(p\!-\!3)(p\!-\!2)(p\!-\!1)p}\!-\!a^{-(p-3)}\!\left(\!\frac{1}{9(p\!-\!3)}
\!-\!\frac{a^{-1}}{3(p\!-\!2)}\!+\!\frac{a^{-2}}{3(p\!-\!1)}\!-\!\frac{a^{-3}}{9\,p}\right)\!\!\!\Bigg\}\!\Bigg]\nonumber\\
&&\hspace{-0.3cm}+\frac{k^4}{H^2a^4}
\!\Bigg[\Big(\frac{\pi^2}{420}\!-\!\frac{548167}{12348000}\Big)a^4
\!+\!\Big(\frac{9403}{54000}\!-\!\frac{\pi^2}{90}\Big)a^3\!+\!\Big(\frac{\pi^2}{45}
\!-\!\frac{8213}{27000}\Big)a^2\!+\!\Big(\frac{425}{864}\!-\!\frac{\pi^2}{36}\Big)a\nonumber\\
&&\hspace{-0.3cm}-\frac{\ln^3(a)}{72}\!-\!\frac{5\ln^2(a)}{96}
\!-\!\Big(\frac{\pi^2}{144}\!+\!\frac{127}{1152}\Big)\ln(a)\!+\!\frac{37733}{564480}\!-\!\frac{467\pi^2}{15120}
\!+\!\frac{\zeta(3)}{12}\!-\!\Big(\frac{857}{13440}\!+\!\frac{\pi^2}{720}\Big)a^{-1}\nonumber\\
&&\hspace{-0.3cm}+\Big(\frac{\pi^2}{1440}
\!+\!\frac{24167}{403200}\Big)a^{-2}\!-\!\Big(\frac{\pi^2}{15120}
\!+\!\frac{65183}{1814400}\Big)a^{-3}\!+\!\frac{11a^{-4}}{1152}\!+\!\frac{a^{-5}}{10080}
\!-\!\frac{13a^{-6}}{12096}\!+\!\frac{11a^{-7}}{70560}\nonumber\\
&&\hspace{-0.3cm}-\frac{1}{840}\!\sum_{n=3}^\infty
\frac{(455\,n^5\!-\!875\,n^4+3787\,n^3-11109\,n^2-8538\,n-2392)a^{-(n+1)}}
{(n\!-\!2)(n\!-\!1)n(n\!+\!1)^3(n\!+\!2)(n\!+\!3)(n\!+\!4)(n\!+\!5)}\nonumber\\
&&\hspace{-0.3cm}-\!\!\sum_{p=5}^{\infty}\!\sum_{n=1}^{p-1}
\!\frac{1}{n(p\!-\!n)}\Bigg\{\!\frac{p^4\!+\!6p^3\!+\!19p^2\!+\!46p\!+\!24}
{4(p\!-\!4)(p\!-\!3)(p\!-\!2)(p\!-\!1)p(p\!+\!1)(p\!+\!2)(p\!+\!3)}\!-\!a^{-(p-4)}
\Big(\frac{2}{35(p\!-\!4)}\nonumber\\
&&\hspace{-0.3cm}-\frac{a^{-1}}{5(p\!-\!3)}\!+\!\frac{4a^{-2}}{15(p\!-\!2)}\!-\!\frac{a^{-3}}{6(p\!-\!1)}
\!+\!\frac{a^{-4}}{24p}\!+\!\frac{a^{-5}}{120(p\!+\!1)}
\!-\!\frac{a^{-6}}{120(p\!+\!2)}\!+\!\frac{a^{-7}}{840(p\!+\!3)}\Big)\!\!\Bigg\}\!\Bigg]
\!\!+\!\mathcal{O}\Big(\frac{ik^5}{H^3}\Big) \!\Bigg\}\;
.\label{I5exp}\nonumber\\\ee Remaining integral
$\mathcal{I}_6(\eta, k)$ involves trivial delta function
integrations which yield\be&&\hspace{-1.4cm}\mathcal{I}_6(\eta,
k)\!=\! \frac{-1}{2^7\pi^4}
\Bigg\{\!\Bigg[\!\ln(a)\frac{\left(\dd_0^2\!+\!k^2\right)}{12\,a^2}
\!+\!\Big(\!2\ln(a)\!+\!1\!\Big)H\frac{\dd_0}{12\,a}\Bigg]\!\!-\!H^2\Bigg[\frac{4\ln^3(a)}{9}
\!+\!\frac{23\ln^2(a)}{18}\nonumber\\
&&\hspace{-1.2cm}+\Big(\frac{2\pi^2}{9}\!+\!3\ln\!\Big(\!\frac{2\mu}{H}\Big)\!-\!\frac{13}{3}\Big)\!\ln(a)
\!+\!\frac{3\,a^{-2}}{4}\!+\!\frac{8\,a^{-3}}{81}
\!+\!\sum_{n=4}^{\infty}\frac{(n\!-\!3)(n\!+\!2)\,a^{-(n+1)}}{(n\!-\!2)(n\!+\!1)^3}\Bigg]\!\Bigg\}u(\eta,k)\;
.\ee Acting the derivatives upon the tree-order mode function
$u(\eta,k)$, we find\be &&\hspace{-0.5cm}\mathcal{I}_6(\eta,
k)\!=\frac{1}{2^7\pi^4}\frac{H}{\sqrt{2k^3}}\Bigg\{\!H^2\Bigg[\frac{4\ln^3(a)}{9}
\!+\!\frac{23\ln^2(a)}{18}\!+\!\Big(\frac{2\pi^2}{9}\!+\!3\ln\!\Big(\!\frac{2\mu}{H}\Big)\!-\!\frac{13}{3}\Big)\!\ln(a)
\!+\!\frac{3\,a^{-2}}{4}\nonumber\\
&&\hspace{-0.4cm}+\frac{8\,a^{-3}}{81}\!+\!\!\sum_{n=4}^{\infty}\frac{(n\!-\!3)(n\!+\!2)\,a^{-(n+1)}}{(n\!-\!2)(n\!+\!1)^3}\Bigg]
\!\!+\!\frac{k^2}{a^2}\Bigg[\frac{2\ln^3(a)}{9}
\!+\!\frac{23\ln^2(a)}{36}\!+\!\Big(\frac{\pi^2}{9}\!+\!\frac{3}{2}\ln\!\Big(\!\frac{2\mu}{H}\Big)
\!-\!\frac{13}{6}\Big)\!\ln(a)\nonumber\\
&&\hspace{-0.4cm}+\frac{1}{12}\!+\!\frac{3\,a^{-2}}{8}\!+\!\frac{4\,a^{-3}}{81}
\!+\!\frac{1}{2}\!\sum_{n=4}^{\infty}\frac{(n\!-\!3)(n\!+\!2)\,a^{-(n+1)}}{(n\!-\!2)(n\!+\!1)^3}\Bigg]
\!+\!\frac{ik^3}{Ha^3}\Bigg[\frac{4\ln^3(a)}{27}
\!+\!\frac{23\ln^2(a)}{54}\nonumber\\
&&\hspace{-0.4cm}+\Big(\frac{2\pi^2}{27}\!+\!\ln\!\Big(\!\frac{2\mu}{H}\Big)\!-\!\frac{13}{9}\Big)\!\ln(a)
\!+\!\frac{1}{12}\!+\!\frac{a^{-2}}{4}\!+\!\frac{8\,a^{-3}}{243}
+\!\frac{1}{3}\!\sum_{n=4}^{\infty}\frac{(n\!-\!3)(n\!+\!2)\,a^{-(n+1)}}{(n\!-\!2)(n\!+\!1)^3}\Bigg]\nonumber\\
&&\hspace{-0.4cm}-\frac{k^4}{H^2a^4}\Bigg[\frac{\ln^3(a)}{18}
\!+\!\frac{23\ln^2(a)}{144}\!+\!\Big(\frac{\pi^2}{36}\!+\!\frac{3}{8}\ln\!\Big(\!\frac{2\mu}{H}\Big)
\!-\!\frac{13}{24}\Big)\!\ln(a)
\!+\!\frac{1}{24}\!+\!\frac{3\,a^{-2}}{32}\!+\!\frac{a^{-3}}{81}\nonumber\\
&&\hspace{6.2cm}+\frac{1}{8}\!\sum_{n=4}^{\infty}
\frac{(n\!-\!3)(n\!+\!2)\,a^{-(n+1)}}{(n\!-\!2)(n\!+\!1)^3}\Bigg]\!\!+\!\mathcal{O}\Big(\frac{ik^5}{H^3}\Big)\!\Bigg\}\;
.\label{I6exp}\ee In the next appendix, we sum up $\mathcal{I}_1(\eta, k)$ given
in Eq.~(\ref{I1exp}) and the five integrals we evaluated in this
appendix to calculate source term~(\ref{STermIntegroDE}) of
integro-differential equation~(\ref{diffeqmainPHI2}).

\section{Calculating the source term $\mathcal{S}(\eta, k)$}
\label{App:S} In Eq.~(\ref{STermIntegroDE}) we defined the
non-homogeneous part of the integro-differential equation for the
two-loop correction $\Phi_2(\eta, k)$ to the scalar mode function,
as the source term\beeq \mathcal{S}(\eta,
k)\equiv-\frac{H^2}{8\pi^2}\ln(a(\eta))\Phi_1(\eta,
k)\!+\!\sum_{n=1}^6 \mathcal{I}_n(\eta,k) \;
.\label{totalSOURCE}\eneq The one-loop correction $\Phi_1(\eta,
k)$ is given in Eqs.~(\ref{exactPhi1}) and (\ref{power1}). The
integrals $\{\mathcal{I}_n(\eta, k):n=1,2,\ldots,6\}$ are given in
Eqs.~(\ref{I1exp}), (\ref{I2exp}), (\ref{I3exp}), (\ref{I4exp}),
(\ref{I5exp}), (\ref{I6exp}), respectively. Inserting these
results into Eq.~(\ref{totalSOURCE}) yields
\be&&\hspace{-0.7cm}\mathcal{S}(\eta,k)=\frac{1}{2^{6}
\,\pi^4}\frac{H}{\sqrt{2k^3}}
\Bigg\{\!H^2\!\Bigg[\frac{11\ln^3(a)}{18}\!+\!\frac{19\ln^2(a)}{36}
\!+\!\Big(\frac{2\pi^2}{9}\!-\!\frac{559}{216}
\!+\!\frac{3\ln\!\left(\frac{2\mu}{H}\right)}{2}\!\Big)\!\ln(a)\nonumber\\
&&\hspace{-0.7cm}+\frac{\ln^2\!\left(\frac{2\mu}{H}\right)}{2}
\!-\!\frac{5\ln\!\left(\frac{2\mu}{H}\right)}{6} \!+\!\frac{5}{12}
\!+\!\frac{13\pi^2}{108}\!-\!\frac{4\zeta(3)}{3}\!+\!\frac{7a^{-2}}{12}
\!-\!\Big(\frac{\ln(a)}{27}\!+\!\frac{\pi^2}{27}\!-\!\frac{109}{648}
\Big)a^{-3}\nonumber\\
&&\hspace{-0.7cm}+\!\sum_{n=3}^\infty\frac{\left(n^7\!-\!2n^6\!-\!8n^5\!+\!46n^4\!+\!19n^3\!-\!188n^2
\!+\!196n\!+\!64\right)a^{-(n+1)}}
{24(n\!-\!2)(n\!-\!1)n(n\!+\!1)^3}\!+\!2\!\sum_{p=2}^{\infty}\!\sum_{n=1}^{p-1}\!\frac{1}{n(p\!-\!n)}\nonumber\\
&&\hspace{-0.7cm}\times\Bigg\{\!\frac{1}{p(p\!+\!3)}
\!-\!\frac{a^{-p}}{4}\!\left[\!1\!-\!p\left(1\!-\!a^{-1}\right)
\!+\!\frac{4}{3}\Big(\frac{1}{p}\!-\!\frac{a^{-3}}{p\!+\!3}\Big)\!\right]\!\!\Bigg\}\!\Bigg]\!
\!+\!k^2\!\Bigg[\frac{\ln(a)}{30}\!+\!\frac{13283}{27000}\!-\!\frac{\pi^2}{45}\!+\!a^{-2}\Big[\frac{11\!\ln^3(a)}{36}\nonumber\\
\nonumber\\
&&\hspace{-0.7cm}-\frac{29\ln^2(a)}{72}
\!+\!\Big(\frac{\pi^2}{9}\!-\!\frac{151}{432}
\!+\!\frac{3\ln\!\left(\frac{2\mu}{H}\right)}{4}\!\Big)\!\ln(a)\!+\!\frac{\ln^2\!\left(\frac{2\mu}{H}\right)}{4}
\!-\!\frac{5\ln\!\left(\frac{2\mu}{H}\right)}{12}\!-\!\frac{347}{1350}\!+\!\frac{77\pi^2}{1080}\!-\!\frac{2\zeta(3)}{3}\Big]\nonumber\\
&&\hspace{-0.7cm}
-\Big(\frac{\ln(a)}{3}\!+\!\frac{\pi^2}{9}\!-\!\frac{59}{40}
\Big)a^{-3}\!-\!\frac{301a^{-4}}{360}\!+\!\Big(\frac{\ln(a)}{270}\!+\!\frac{\pi^2}{270}\!+\!\frac{5207}{6480}\Big)a^{-5}
\!+\!\frac{2671a^{-6}}{21600}\!+\!\frac{909a^{-7}}{7000}\nonumber\\
&&\hspace{-0.7cm}+\!\sum_{n=5}^\infty\!\frac{\left(5n^9\!+\!35n^8\!+\!40n^7\!+\!10n^6\!+\!645n^5\!-\!905n^4\!-\!6170n^3\!-\!1860n^2
\!+\!6152n\!+\!512\right)\!a^{-(n+3)}}
{240(n\!-\!2)(n\!-\!1)n(n\!+\!1)^3(n\!+\!2)(n\!+\!3)}\nonumber\\
&&\hspace{-0.7cm}-\sum_{p=3}^{\infty}\!\sum_{n=1}^{p-1}\!\frac{1}{n(p\!-\!n)}\Bigg\{\!\frac{2a^{-2}}{(p\!-\!2)p(p\!+\!1)(p\!+\!3)}
\!-\!a^{-p}\Bigg[\!\frac{(5p^2\!-\!5p\!-\!8)}{30(p\!-\!2)}\!-\!\frac{(3p^2\!+\!9p\!+\!4)a^{-2}}{12p}\nonumber\\
&&\hspace{-0.7cm}+\frac{(p^2\!+\!5p\!+\!8)a^{-3}}{12(p\!+\!1)}\!-\!\frac{a^{-5}}{15(p\!+\!3)}\Big)\!\Bigg]\!\Bigg\}\!\Bigg]
\!\!+\!\frac{ik^3}{H}\!\Bigg[\frac{\ln(a)}{81}\!+\!\frac{155}{648}\!-\!\frac{\pi^2}{81}
\!+\!\frac{5a^{-2}}{4}\!-\!\frac{\ln^3(a)}{18a^3}\!-\!\frac{7\ln^2(a)}{108a^3}\nonumber\\
&&\hspace{-0.7cm}+\Big(\frac{2285}{4536}
\!+\!\frac{\ln\!\!\left(\frac{2\mu}{H}\right)}{2}\!\Big)\!\frac{\ln(a)}{a^3}\!+\!\Big(\frac{\ln^2\!\!\left(\frac{2\mu}{H}\right)}{6}
\!-\!\frac{5\!\ln\!\left(\frac{2\mu}{H}\right)}{18}\!+\!\frac{4\zeta(3)}{9}\!-\!\frac{13\pi^2}{108}
\!-\!\frac{12023}{13608}\Big)a^{-3}\!-\!\frac{185}{108a^4}\!+\!\frac{103}{270a^{5}}\nonumber\\
&&\hspace{-0.7cm}+\frac{989}{4860a^{6}}\!+\!\!\sum_{n=3}^\infty\!
\frac{(n^8\!+\!14n^7\!+\!68n^6\!+\!146n^5\!+\!239n^4\!+\!80n^3\!-\!2116n^2\!-\!5056n
\!-\!2720)a^{-(n+4)}}{72(n\!-\!2)(n\!+\!1)^3(n\!+\!2)(n\!+\!3)(n\!+\!4)}\nonumber\\
&&\hspace{-0.7cm}-\!\sum_{p=4}^{\infty}\!\sum_{n=1}^{p-1}\!\frac{1}{n(p\!-\!n)}
\Bigg\{\!\frac{(p\!+\!4)a^{-3}}{3(p\!-\!3)(p\!-\!2)p}\!-\!\frac{a^{-p}}{18}\Bigg(\!\frac{3p^2\!-\!3p\!-\!4}{p\!-\!3}
\!-\!\frac{3p(p\!+\!1)a^{-1}}{p\!-\!2}\!+\!\frac{4a^{-3}}{p}\!\Bigg)\!\Bigg\}\Bigg]\nonumber\\
&&\hspace{-0.7cm}-\frac{k^4}{H^2}
\!\Bigg[\frac{\ln(a)}{280}\!-\!\frac{\pi^2}{210}\!+\!\frac{548167}{6174000}
\!-\!\frac{\ln(a)}{60a^2}\!+\!\Big(\frac{\pi^2}{90}
\!+\!\frac{4477}{54000}\Big)a^{-2}\!-\!\frac{a^{-3}}{6}\!+\!\frac{11\ln^3(a)}{144a^4}\!+\!\frac{19\ln^2(a)}{288a^4}\nonumber\\
&&\hspace{-0.7cm}-\Big(\!\frac{199}{1728}\!-\!\frac{\pi^2}{36}
\!-\!\frac{3\ln\!\left(\frac{2\mu}{H}\right)}{16}\!\Big)\!\frac{\ln(a)}{a^4}\!+\!\Big(\frac{88663}{235200}
\!+\!\frac{1343\pi^2}{30240}\!+\!\frac{\ln^2\!\left(\frac{2\mu}{H}\right)}{16}
\!-\!\frac{5\ln\!\left(\frac{2\mu}{H}\right)}{48}\!-\!\frac{\zeta(3)}{6}\Big)a^{-4}\nonumber\\
&&\hspace{-0.7cm}-\frac{\ln(a)}{30a^5}\!-\!\Big(\frac{\pi^2}{90}
\!+\!\frac{719}{560}\Big)a^{-5}\!+\!\frac{51577}{50400a^{6}}\!+\!\frac{\ln(a)}{7560a^7}\!-\!\Big(\frac{113849}{907200}
\!-\!\frac{\pi^2}{7560}\Big)a^{-7}\!-\!\frac{7a^{-8}}{40}\!+\!\frac{1297a^{-9}}{25200}\nonumber\\
&&\hspace{-0.7cm}-\frac{a^{-10}}{2520}
\!-\!\frac{11a^{-11}}{35280}\!+\!\sum_{n=3}^\infty\!\frac{\left(455n^5\!-\!875n^4\!+\!3787n^3
\!-\!11109n^2\!-\!8538n\!-\!2392\right)\!a^{-(n+5)}}{420(n\!-\!2)(n\!-\!1)n(n\!+\!1)^3(n\!+\!2)(n\!+\!3)(n\!+\!4)(n\!+\!5)}\nonumber\\
&&\hspace{-0.7cm}+\!\sum_{n=3}^\infty\!\frac{\left(5n^8\!+\!80n^7\!+\!450n^6\!+\!1160n^5\!+\!1805n^4\!+\!760n^3
\!+\!252n^2\!-\!5664n\!-\!1856\right)\!a^{-(n+5)}}
{960(n\!-\!1)n(n\!+\!1)^2(n\!+\!2)(n\!+\!3)(n\!+\!4)}\nonumber\\
&&\hspace{-0.7cm}+\!\sum_{n=3}^\infty\frac{(n\!-\!3)(n\!+\!2)a^{-(n+5)}}{16(n\!-\!2)(n\!+\!1)^3}
-\!\!\sum_{p=5}^{\infty}\!\sum_{n=1}^{p-1}\!\frac{1}{n(p\!-\!n)}\Bigg\{\!\frac{(p^4\!+\!8p^3\!+\!21p^2
\!+\!22p\!+\!24)a^{-4}}{4(p\!-\!4)(p\!-\!3)(p\!-\!2)p(p+\!1)(p\!+\!3)}\nonumber\\
&&\hspace{-0.7cm}-a^{-p}\!\Bigg[\frac{7p^2\!-\!7p\!-\!8}{70(p\!-\!4)}\!-\!
\frac{p(p\!+\!1)a^{-1}}{6(p\!-\!3)}\!+\!\frac{(5p^2\!+\!15p+\!8)a^{-2}}{60(p\!-\!2)}\!-\!\frac{(p^2\!+\!7p\!+\!4)a^{-4}}{48p}\nonumber\\
&&\hspace{5cm}+\frac{(p^2\!+\!9p+\!16)a^{-5}}{240(p+\!1)}\!-\!
\frac{a^{-7}}{420(p+\!3)}\Bigg]\!\Bigg\}\!\Bigg]\!+\!\mathcal{O}\Big(\frac{ik^5}{H^3}\Big)
\!\Bigg\}\; .\label{inhomogen}\ee When the source
term~(\ref{inhomogen}) is integrated against the Green's
function~(\ref{Green}) in Eq.~(\ref{Phi2intso}) gives the two-loop
correction $\Phi_2(\eta, k)$ of Eq.~(\ref{Phi2}).

\section{Evaluating the functional $\mathcal{W}(\alpha)$ }
\label{App:w} In calculating the third integral $I_3(\eta, k)$ of
the source term $\mathcal{S}(\eta, k)$ in Appendix~B we defined
the functional~$\mathcal{W}(\alpha)$ in Eq.~(\ref{Walpha}) as\be
\mathcal{W}(\alpha)\!&=&\!\!\int_0^1 \!\!dz z \sin(\alpha
z)\ln^2\!\left(\!\frac{1\!-\!z^2}{4}\!\right)\\\!&=&\!\!\sum_{n=0}^\infty
\frac{(-1)^n\alpha^{2n+1}}{(2n\!+\!1)!}\!\int_0^1 \!\!dz
z^{2n+2}\ln^2\!\left(\!\frac{1\!-\!z^2}{4}\!\right)\;
.\label{newint}\ee In this appendix we calculate the
$\mathcal{W}(\alpha)$. To evaluate the integral in
Eq.~(\ref{newint}), let us first consider the following integral
\beeq \int_0^1 \!\!dz z^{2n+2}\left(\!\frac{1-z^2}{4}\!\right)^b\;
.\eneq Making the change of variable $z^2\equiv t$ yields \be
\int_0^1\!\!dz
z^{2n+2}\left(\!\frac{1\!-\!z^2}{4}\!\right)^b\!&=&\!\frac{1}{2^{2b+1}}\!\int_0^1
\!\!dt\,
t^{n+\frac{1}{2}}\,(1\!-t)^b=\frac{1}{2^{2b+1}}\mathcal{B}\Big(\!n\!+\!\frac{3}{2},
b+\!1\!\Big)\nonumber\\\!&=&\!\frac{\Gamma(1\!+\!
b)\Gamma(\frac{3}{2}\!+\!n)}{2^{2b+1}\Gamma(\frac{5}{2}\!+\!b+\!
n)}\; , \label{alp}\ee which is valid for arbitrary $b$ and $n$ as
long as $Re[b]>-1$ and $Re[n]
> -\frac{3}{2}$. $\mathcal{B}$ denotes the Beta function. Differentiating
the both sides of Eq.~(\ref{alp}) and taking $b\rightarrow0$, we
obtain the desired integral in Eq.~(\ref{newint}) as \be
\hspace{-1cm}\int_0^1\!\!dz\,
z^{2n+2}\ln^2\!\!\left(\!\frac{1\!-\!z^2}{4}\!\right)\!\!
&=&\!\!\frac{\Gamma(\frac{3}{2}\!+\!n)}{2\Gamma(\frac{5}{2}
\!+\!n)}\Bigg\{\!4\ln(2)\!\Big[\ln(2)\!+\!\mathcal{H}_{\scriptscriptstyle\frac{3}{2}+n}\Big]
\!-\!\psi'\Big(\frac{5}{2}\!+\!n\!\Big)
\!+\!\mathcal{H}^2_{\scriptscriptstyle\frac{3}{2}+n}\!+\!\frac{\pi^2}{6}\!
\Bigg\}\; ,\label{int}\ee where the Digamma Function $\psi(z)$ is
the logarithmic derivative of the Gamma Function,
$\psi(z)\!=\!\frac{\Gamma'(z)}{\Gamma(z)}$, and the Harmonic
Number Function
$\mathcal{H}_{\scriptscriptstyle\frac{3}{2}+n}\!=\!\gamma\!+\!\psi\left(\frac{5}{2}\!+\!n\right)$.
Prime denotes the derivative with respect to the argument. Note
that, if we had done the calculation of the third integral
$I_3(\eta, k)$ in a way that the factor $1/4$ in the argument of
the logarithm had not been included in definition~(\ref{Walpha})
of the $\mathcal{W}(\alpha)$, the first term on the right hand
side of Eq.~(\ref{int}) would not have been present there.
Although Eq.~(\ref{int}) would look compact, that result, however,
would yield complicated terms for each integer $n$.
Equation~(\ref{int}), on the other hand, yields simpler terms for
$n=0$, $n=1$ and $n=2$, \be \int_0^1 \!dz
z^2\ln^2\!\left(\!\frac{1\!-\!z^2}{4}\!\right)\!&=&\!\frac{104}{27}\!-\!\frac{\pi^2}{9}\; ,\label{intforn0}\\
\int_0^1 \!dz
z^4\ln^2\!\left(\!\frac{1\!-\!z^2}{4}\!\right)\!&=&\!\frac{3152}{1125}\!-\!\frac{\pi^2}{15}\; ,\label{intforn1}\\
\int_0^1 \!dz
z^6\ln^2\!\left(\!\frac{1\!-\!z^2}{4}\!\right)\!&=&\!\frac{175568}{77175}\!-\!\frac{\pi^2}{21}\;
,\label{intforn2}\ee respectively. (If we had odd powers of $z$ in
Eq.~(\ref{int}), then the result of the integral for each integer
$n$ would be simpler without the inclusion of the factor $1/4$ in
the argument of the logarithm, in the definition of the
$\mathcal{W}(\alpha)$.) Using
Eqs.~(\ref{intforn0})-(\ref{intforn2}) in Eq.~(\ref{newint}) we
obtain\be
\mathcal{W}(\alpha)\!=\!\left(\!\frac{104}{27}\!-\!\frac{\pi^2}{9}\!\right)\!\alpha
\!-\!\left(\!\frac{1576}{3375}\!-\!\frac{\pi^2}{90}\!\right)\!\alpha^3
\!+\!\left(\!\frac{21946}{1157625}\!-\!\frac{\pi^2}{2520}\!\right)\!\a^5\!+\!\mathcal{O}(\alpha^7)\;
.\ee This functional also appears in the evaluations of the
integrals $I_4(\eta, k)$ and $I_5(\eta, k)$. See
Eqs.~(\ref{WalphaI4}) and (\ref{WalphaI5}), respectively.
\end{appendix}

\begin{acknowledgements}
I am grateful to Richard P. Woodard for stimulating discussions.
\end{acknowledgements}


\begin{thebibliography}{99}

\bibitem{RWo1} R. P. Woodard, ``Quantum Effects during Inflation,''
in {\it Quantum Field Theory under the Influence of External
Conditions} (Rinton Press, Princeton, 2004) ed.  K. A. Milton, pp.
325-330, astro-ph/0310757.


\bibitem{MuCh} V. F. Mukhanov and G. V. Chibisov, JETP Lett. {\bf 33}
(1981) 532.

\bibitem{ASt1} A. A. Starobinski\u{\i}, JETP Lett. {\bf 30} (1979) 682.


\bibitem{OnWo1} V. K. Onemli and R. P. Woodard, Class. Quant. Grav. {\bf 19}
(2002) 4607.

\bibitem{OnWo2} V. K. Onemli and R. P. Woodard, Phys. Rev. {\bf D70} (2004) 107301.

\bibitem{KaOnWo} E. O Kahya, V. K. Onemli and R. P. Woodard, Phys. Rev. {\bf D81} (2010) 023508.

\bibitem{BrOnWo} T. Brunier, V. K. Onemli and R. P. Woodard, Class. Quant.
Grav. {\bf 22} (2005) 59.

\bibitem{KaOn} E. O. Kahya and V. K. Onemli, Phys. Rev. {\bf D76} (2007) 043512.

\bibitem{PrWo1} T. Prokopec and R. P. Woodard, JHEP {\bf 0310} (2003) 059.

\bibitem{DuWo} L. D. Duffy and R. P. Woodard, Phys. Rev. {\bf D72} (2005)
024023.

\bibitem{MiWo1} S. P. Miao and R. P. Woodard, Phys. Rev. {\bf D74} (2006)
044019.

\bibitem{PrToWo} T. Prokopec, O. Tornkvist and R. P. Woodard, Phys. Rev.
Lett. {\bf 89} (2002) 101301; Ann. Phys. {\bf 303} (2003) 251; T.
Prokopec and R. P. Woodard, Ann. Phys. {\bf 312} (2004) 1.

\bibitem{KaWo1} E. O. Kahya and R. P. Woodard, Phys. Rev. {\bf D72} (2005)
104001; Phys. Rev. {\bf D74} (2006) 084012.

\bibitem{MiWo2} S. P. Miao and R. P. Woodard, Class. Quant. Grav. {\bf 23}
(2006) 1721; Phys. Rev. {\bf D74} (2006) 024021.

\bibitem{KaWo2} E. O. Kahya and R. P. Woodard, Phys. Rev. {\bf D77} (2008)
084012.


\bibitem{Sc} J. Schwinger, J. Math. Phys. {\bf 2} (1961) 407;
K. T. Mahanthappa, Phys. Rev. {\bf 126} (1962) 329; P. M. Bakshi
and K. T. Mahanthappa, J. Math. Phys. {\bf 4} (1963) 1; L. V.
Keldysh, Sov. Phys. JETP {\bf 20} (1965) 1018.

\bibitem{Jor} R. D. Jordan, Phys. Rev. {\bf D33} (1986) 444;
K. C. Chou, Z. B. Su, B. L. Hao and L. Yu, Phys. Rept.
{\bf 118} (1985) 1; E. Calzetta and B. L. Hu, Phys. Rev. {\bf D35}
(1987) 495.

\bibitem{MiPa} S. P. Miao and S. Park, arXiv:1306.4126.

\bibitem{LoopRefs} J. Maldacena, JHEP {\bf 0305} (2003) 013;
S. Weinberg, Phys. Rev. {\bf D72} (2005) 043514; Phys. Rev. {\bf
D74} (2006) 023508; D. Boyanovsky, H. J. de Vega and N. G.
Sanchez, Nucl. Phys. {\bf B747} (2006) 25; Phys. Rev. {\bf D72}
(2005) 103006; M. Sloth, Nucl. Phys. {\bf B748} (2006) 149; Nucl.
Phys. {\bf B775} (2007) 78; D. Seery, J. E. Lidsey and M. S.
Sloth, JCAP {\bf 0701} (2007) 027; M. van der Meulen and J. Smit,
JCAP {\bf 0711} (2007) 023; D. H. Lyth, JCAP {\bf 0712} (2007)
016; D. Seery, JCAP {\bf 0711} (2007) 025; JCAP {\bf 0802} (2008)
006; JCAP {\bf 0905} (2009) 021; Class. Quant. Grav. {\bf 27}
(2010) 124005; N. Bartolo, S. Matarrese, M. Pietroni, A. Riotto
and D. Seery, JCAP {\bf 0801} (2008) 015; Y. Urakawa and K. I.
Maeda, Phys. Rev. {\bf D78} (2008) 064004; A. Riotto and M. Sloth,
JCAP {\bf 0804} (2008) 030; JCAP {\bf 1110} (2011) 003; K.
Enqvist, S. Nurmi, D. Podolsky and G. I. Rigopoulos, JCAP {\bf
0804} (2008) 025; P. Adshead, R. Easther and E. A. Lim, Phys. Rev.
{\bf D79} (2009) 063504; E. O. Kahya, V. K. Onemli and R. P.
Woodard, Phys. Lett. {\bf B694} (2010) 101; Y. Urakawa and T.
Tanaka, Prog. Theor. Phys. {\bf 122} (2009) 779; Prog. Theor.
Phys. {\bf 122} (2010) 1207; Phys. Rev. {\bf D82} (2010) 121301;
Prog. Theor. Phys. {\bf 125} (2011) 1067; JCAP {\bf 1105} (2011)
014; Prog. Theor. Exp. Phys. {\bf 2013} (2013) 6, 063E02; Prog.
Theor. Exp. Phys. {\bf 2013} (2013) 8 083E01; Y. Urakawa, Prog.
Theor. Phys. {\bf 126} (2011) 961; S. B. Giddings and M. S. Sloth,
JCAP {\bf 1007} (2010) 015; JCAP {\bf 1101} (2011) 023; Phys. Rev.
{\bf D84} (2011) 063528; Phys. Rev. {\bf D86} (2012) 083538; J.
Kumar, L. Leblond and A. Rajaraman, JCAP {\bf 1004} (2010) 024; C.
P. Burgess, R. Holman, L. Leblond and S. Shandera, JCAP {\bf 1003}
(2010) 033; L. Senatore and M. Zaldarriaga, JHEP {\bf 1012} (2010)
008; M. Gerstenlauer, A. Hebecker and G. Tasinato, JCAP {\bf 1106}
(2011) 021; W. Xue, X. Gao and R. Brandenberger, JCAP {\bf 1206}
(2012) 035; G. L. Pimental, L. Senatore and M. Zaldarriaga, JHEP
{\bf 1207} (2012) 166; S. P. Miao and R. P. Woodard, JCAP {\bf
1207} (2012) 008; M. G. Romania, N. C. Tsamis and R. P. Woodard,
JCAP {\bf 1208} (2012) 029; D. Boyanovsky, Phys. Rev. D {\bf 85}
(2012) 123525; Phys. Rev. D {\bf 86} (2012) 023509; K. Feng, Y.-F.
Cai and Y.-S. Piao, Phys. Rev. D {\bf 86} (2012) 103515; A. Kaya,
Phys. Rev. D {\bf 86} (2012) 123511; arXiv:1306.3236; R.
Brandenberger and J. Martin, Class. Quant. Grav. {\bf 30} (2013)
113001; Y. Korai and T. Tanaka, Phys. Rev. D {\bf 87} (2013)
024013; E.T. Akhmedov, Int. J. Mod. Phys. {\bf D23} (2014)
1430001; K. Larjo and D. A. Lowe, Phys. Rev. D {\bf 87} (2013)
083506; J. Serreau and R. Parentani, Phys. Rev. D {\bf 87} (2013)
085012; J. Serreau, arXiv:1302.6365; L. Lello, D. Boyanovsky and
R. Holman arXiv:1307.4066.

\bibitem{BeSh} F. L. Bezrukov and M. E. Shaposhnikov, Phys. Lett. {\bf B659} (2008) 703.

\end{thebibliography}
\end{document}